\documentclass[12pt]{article}
\usepackage{arxiv}
\usepackage[colorlinks=true,citecolor=blue]{hyperref}
\usepackage{graphicx}
\usepackage{amsmath,amssymb}
\usepackage{amsthm}
\usepackage{color}
\usepackage{dsfont}
\usepackage{bm}
\usepackage{mathpazo,times}
\usepackage{enumitem}  

\theoremstyle{definition}
\newtheorem{definition}{Definition}
\DeclareMathOperator{\sech}{sech}
\DeclareMathOperator{\sgn}{sgn}

\newcommand{\ub}{\bar{u}}

\renewcommand{\shorttitle}{Oblique soliton and mean flow interactions}

\advance\textheight-12mm
\advance\voffset-2mm

\def\Wr{\mathop{\mathrm{Wr}}\nolimits}

\begin{document}

\title{Oblique interactions between solitons and mean flows in the Kadomtsev-Petviashvili equation} 
\author{Samuel J. Ryskamp\thanks{ Department of Applied Mathematics, University 
of Colorado, Boulder, CO (samuel.ryskamp@colorado.edu)} \and Mark A. Hoefer\thanks{ Department of Applied Mathematics, University 
of Colorado, Boulder, CO} \and Gino Biondini\thanks{Department of Mathematics and Department of Physics, State University of New York,  Buffalo, NY}}

\maketitle
\begin{abstract}
   The interaction of an oblique line soliton with a one-dimensional dynamic mean flow is analyzed using the Kadomtsev-Petviashvili II (KPII) equation. Building upon previous studies that examined the transmission or trapping of a soliton by a slowly varying rarefaction or oscillatory dispersive shock wave in one space and one time dimension, this paper allows for the incident soliton to approach the changing mean flow at a nonzero oblique angle. By deriving invariant quantities of the soliton--mean flow modulation equations---a system of three (1+1)-dimensional quasilinear, hyperbolic equations for the soliton and mean flow parameters---and positing the initial configuration as a Riemann problem in the modulation variables, it is possible to derive quantitative predictions regarding the evolution of the line soliton within the mean flow. It is found that the interaction between an oblique soliton and a changing mean flow leads to several novel features not observed in the (1+1)-dimensional reduced problem. Many of these interesting dynamics arise from the unique structure of the modulation equations that are nonstrictly hyperbolic, including a well-defined multivalued solution interpreted as a solution of the (2+1)-dimensional soliton--mean modulation equations, in which the soliton interacts with the mean flow and then wraps around to interact with it again. Finally, it is shown that the oblique interactions between solitons and dispersive shock wave solutions for the mean flow give rise to all three possible types of 2-soliton solutions of the KPII equation.
   The analytical findings are quantitatively supported by direct numerical simulations.
\end{abstract}

\section{Introduction}
\label{sec:intro}
 The study of interactions between rapidly-varying waves and a slowly-varying mean flow is a problem of fundamental importance in fluid mechanics and other media. It arises naturally in applications such as internal water waves \cite{semin_2016,bremer_2014,dasaro_1995}, atmospheric waves \cite{pfeffer_1981,atmosphere_andrews}, and shallow water waves \cite{buhler_jacobson_2001,buhler_barreiro_2007}. Wave--mean flow interactions are especially relevant in geophysical fluid dynamics, as the typical scales of gravity waves are too small to be resolved in large-scale numerical simulations, but these features can still affect the evolution of the system \cite{buhler_mcintyre_1998,buhler_2014}. A related problem of recent interest, especially pertaining to internal waves, is wave-current interactions \cite{constantin_2019,compelli_2019}. Slowly-varying currents are ubiquitous in oceans, rivers, and canals, and they can significantly influence the propagation of a disturbance through a fluid. Wave--mean flow interactions can also be considered in other applications such as nonlinear optical and matter waves \cite{Biondini_soli_trapping_2019,Sprenger_tunneling_2018}.
\par In many of these applications, the study of solitary waves is of particular importance, as these objects move rapidly and can transfer large amounts of energy \cite{constantin_2019}. For example, internal solitary waves---frequently imaged via aerial photography \cite{pacific_solitons_1997,wang_internal_observed}---can have both small-scale impacts on deep water objects \cite{Soomere2011} and large-scale effects on climates and currents \cite{constantin_2019}. Consequently, a number of recent studies have examined the interactions of solitons with a changing background flow in the context of nonlinear, dispersive systems in one space and one time dimension (referred to as (1+1)-dimensional), including the Korteweg-de Vries (KdV) equation \cite{ablowitz_cole_2018,maiden_hoefer_2018}, the focusing nonlinear Schr\"{o}dinger (NLS) equation \cite{Biondini_soli_trapping_2019,Biondini_soli_trapping_2018}, the defocusing NLS equation \cite{Sprenger_tunneling_2018}, and the conduit equation \cite{maiden_hoefer_2018,maiden_hoefer_2016}. The analysis of soliton--mean flow interactions for these long wavelength models provides a foundation for investigating more complex, realistic flows \cite{buhler_mcintyre_1998}.

\par The most fundamental soliton--mean flow problem considers a dynamic mean flow $\ub$ that results from a Riemann problem \cite{Biondini_soli_trapping_2019,ablowitz_cole_2018,maiden_hoefer_2018}, that is, a problem in which the initial configuration consists of a discontinuous jump between two constant values \cite{courant_hilbert_1960,Lax_1973}. Most commonly, solutions include rarefaction waves (RWs) for expansive initial conditions and dispersive shock waves (DSWs) for compressive initial conditions.  When a soliton is normally incident to a RW or DSW, two possible outcomes have been identified \cite{maiden_hoefer_2018}. First, the soliton can pass entirely through the mean flow with some change in parameters, a phenomenon known as soliton transmission or tunnelling. However, if the soliton has insufficient amplitude and velocity to surmount the RW or DSW, the soliton remains trapped within the changing mean flow, termed soliton trapping. Solutions to the Riemann problem form building blocks that can be generalized to other types of mean flows.

\par Multiple-scale analysis is a natural tool for many wave--mean flow problems. Linear wave--mean flow interactions were first studied by a scale separation technique in \cite{whitham,bretherton_1968}, which has been generalized to the nonlinear wave setting using Whitham modulation theory \cite{grimshaw_1984}. This approach utilizes averaged conservation laws or an averaged Lagrangian to approximate nonlinear wave dynamics with a quasilinear, hyperbolic system of equations for the wave parameters that vary on long space and slow time scales \cite{whitham}.  In general, it is possible to calculate adiabatically invariant quantities across a changing mean flow, which can then be used to determine parameters on either side of the RW or DSW. Whitham modulation theory \cite{dsw_review_2016} has proven quite effective at predicting changes in wave parameters through the mean flow \cite{Sprenger_tunneling_2018,maiden_hoefer_2018,congy_el_hoefer_2019}.
\par
\begin{figure}
    \centering
    \includegraphics[scale=.25]{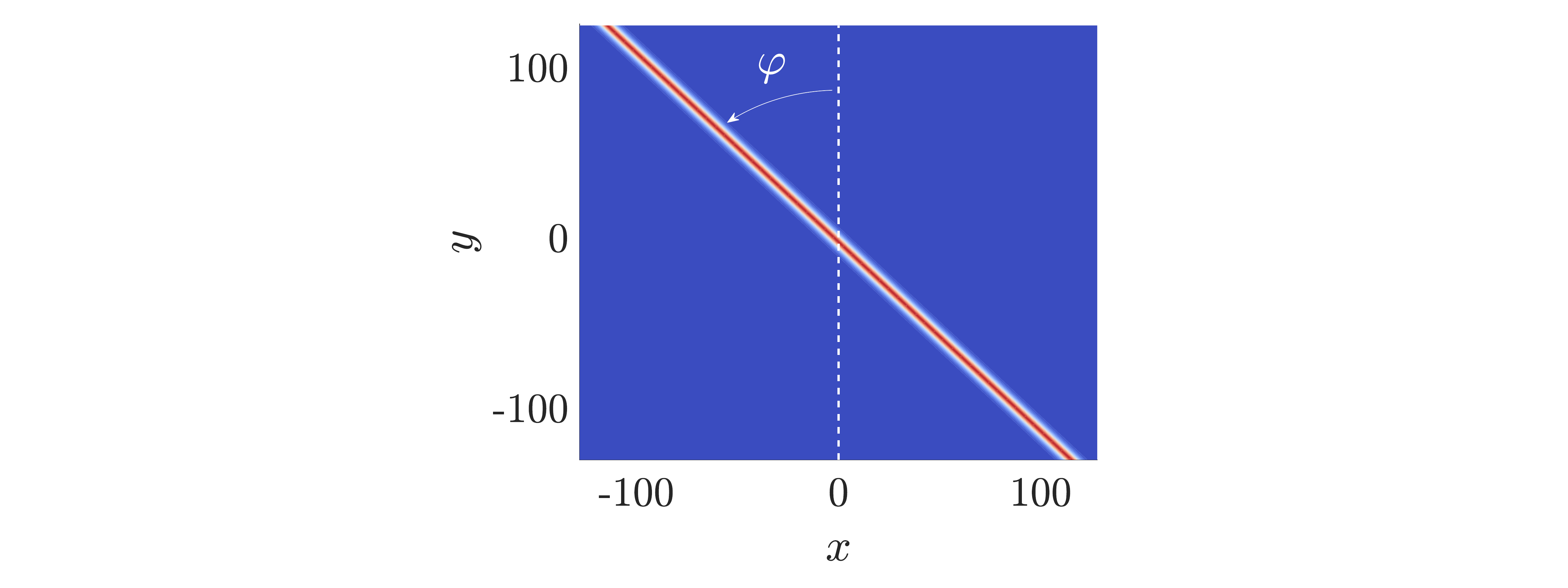}
    \caption{Contour plot of a KP soliton, where $q=\tan{\varphi}$ is a measure of the soliton inclination relative to the $y$-axis.}
    \label{fig:soliton}
\end{figure}

 However, previous studies of soliton--mean flow interactions have only considered governing equations in one space and one time dimension. One important unanswered question is how the transmission or trapping of solitons is affected by a spatial perturbation of the soliton along a direction different from the propagation direction. In this study, we allow a soliton to approach the mean flow at a nonzero incident angle $q$  and examine how this transverse inclination affects the soliton-mean flow interaction. We will do this analysis in the framework of the initial value problem for the Kadomtsev-Petviashvili (KP) equation, originally derived to study the multidimensional stability of KdV solitons by introducing a generalization of the KdV equation \cite{kadomtsev_stability_1970}:
\begin{equation}
    \label{eq:kp}
    (u_t + uu_x + u_{xxx})_x + \beta u_{yy}=0, \quad (x,y) \in \mathds{R}, \quad t>0,
\end{equation}
subject to $u(x,y,0)=u_0(x,y)$, where $\beta = \pm 1$. It is well known that only $\beta=1$ (known as KPII) leads to stable, travelling line soliton solutions, so that is the case we will exclusively consider here. KPII line solitons are a three-parameter family of travelling wave solutions with amplitude $a$ on a background or mean flow $\ub$,
\begin{equation}
    \label{eq:soli}
    u(x,y,t)=\ub + a\sech^2\left(\sqrt{\frac{a}{12}}(x+qy-ct)\right), \quad c = \ub+\frac{a}{3} + q^2,
\end{equation}
where $q=\tan{\varphi}$ is a measure of the transverse inclination of the soliton (see figure~\ref{fig:soliton}) and $c$ is the soliton propagation velocity in the $x$-direction. When $q=0$, \eqref{eq:soli} reduces to the well-known KdV soliton. The goal of this work is to understand and classify the interactions of oblique line solitons \eqref{eq:soli} with one-dimensional mean flows, i.e. the solutions to the KdV Riemann problem. 

\par This problem has both physical and mathematical interest. The KP equation \eqref{eq:kp} has been utilized to model surface water waves \cite{ablowitz_baldwin_2012,ablowitz_evolution_1979,kodama_book}, internal water waves \cite{grimshaw_evolution_1981,grimshaw_internal_2018}, and ion-acoustic waves in plasma \cite{infeld_2001}. The universal character of the KdV equation also translates to the KP equation, and the general nature of the problem means that its principal insights hold true over a wide variety of physical scenarios. Our method for solving this problem---analyzing the $y$-independent KP soliton modulation equations---yields interesting mathematical features as well. For example, this diagonalizable, 3-component, quasilinear system is nonstrictly hyperbolic \cite{biondini2019integrability}, causing some initial conditions to become multivalued. In this scenario, we will need to appeal to the (2+1)-dimensional soliton modulation system to determine regimes of validity for the (1+1)-dimensional multivalued solution.
 \par The addition of an oblique incident angle yields a rich variety of phenomena. Oblique soliton transmission and trapping can both occur either from an initial soliton to the left or the right of the mean flow, unlike one-dimensional solitons which have directionally limited transmission or trapping. The trapping of a soliton starting to the right of a DSW is shown to be closely related to the much-studied effect of line soliton resonance \cite{Miles1977resonant}. Another novel effect is that large soliton amplitude alone is insufficient to guarantee transmission. Line solitons can also experience ``incomplete" transmission, a behaviour not observed in the (1+1)-dimensional reduced problem. This is where the line soliton transmits through the mean flow yet fails to separate from it, even for large times $t$.
 \par The paper is organized as follows. In section~\ref{sec:problem_form}, we introduce the soliton modulation system, its properties, and the Riemann initial conditions examined throughout the rest of this paper. In section~\ref{sec:riem} we look for simple wave solutions to the Riemann problem, where only one Riemann invariant is changing for all $x\in \mathrm{R}$ and $t>0$, and the other two are constant. We expect these self-similar solutions to describe the stable, long-time behaviour of a wide variety of initial and boundary configurations, since simple waves serve as ``attractors" for diverse initial conditions. Next, in sections~\ref{sec:gen} and \ref{sec:spec} we examine one particular class of initial conditions that can give rise to the simple wave solutions found in section~\ref{sec:riem}, as well as other interesting phenomena: a partial soliton encountering step initial conditions in $x$. We conclude this work with some discussion in section~\ref{sec:concl}. \ref{sec:eq_simp_waves} includes a calculation relevant to our analysis.

\par Our analysis is supported by numerical simulations of the KPII equation \eqref{eq:kp} using a Fourier pseudospectral method adapted from \cite{kao_numerical_2012} that allows for outgoing line solitons at the top and bottom of the simulation domain
through use of a windowing function. The numerical scheme is essentially the same as that used in \cite{ryskamp_2020}. To maintain periodicity in $x$, the initial conditions implemented are actually a large box shape -- an upward step near the left side and a corresponding downward step on the right. Simulations are terminated before the edge of the box which is not of interest interferes with the test domain. As in \cite{ryskamp_2020}, step initial conditions are smoothed by using a hyperbolic tangent function to minimize the generation of spurious oscillations that are not described by modulation theory. Additionally, we find that utilizing the windowing function for an initial partial soliton on a background $\ub \neq 0$ leads to numerical instabilities. Consequently, a Galilean transformation is applied so that the partial soliton is initialized on the zero background $\ub=0$. For example, for a partial soliton starting to the right of a RW, we choose $\ub_{\rm R}=0$ and $\ub_{\rm L}=-1$, while for the soliton starting to the left of a RW we set $\ub_{\rm R}=1$ and $\ub_{\rm L}=0$. Most simulations are performed on the spatial domain  $[-1024,1024]\times [-512,512]$ or one similar in size, with spatial and temporal discretisations $\Delta x=\Delta y = 1/2$ and $\Delta t = 10^{-3}$, respectively. Calculations are performed in single precision. Our method and numerical parameter selections are validated in \cite{ryskamp_2020}. In order to quantitatively compare numerical simulation and analytical prediction, we often shift the analytical solution by a relatively small phase shift $x_0$, since phase shifts are a higher-order effect not captured by leading order modulation theory. 
 
\section{Problem formulation and preliminary considerations}
\label{sec:problem_form}

In this section we introduce the soliton modulation system, some of its properties, and the specific initial conditions studied in this paper. We also review features of multi-soliton solutions of the KPII equations that will be relevant later.
\subsection{Modulation system}
The Whitham modulation equations for the KP equation \eqref{eq:kp} were recently derived in \cite{ablowitz_2017_whitham}. The soliton limit of the equations, a (2+1)-dimensional hyperbolic system consisting of four equations, was further analyzed in \cite{biondini2019integrability}. Assuming that the mean flow and soliton modulation parameters only change in the propagation direction of the mean flow, $x$, the modulation equations simplify to a (1+1)-dimensional set of equations for the three parameters in~\eqref{eq:soli}
\begin{equation}
    \label{eq:mf1}
    \begin{bmatrix}
    \ub \\ a \\ q
  \end{bmatrix}_t +
  \begin{bmatrix}
    \ub & 0 & 0 \\
    \frac{2}{3} a & \ub + \frac{a}{3} - q^2 & - \frac{4}{3}
     a q 
    \\
    -q & -\frac{1}{3} q & \ub + \frac{a}{3} - q^2
  \end{bmatrix}
  \begin{bmatrix}
    \ub \\ a \\ q
  \end{bmatrix}_x = 0 .
\end{equation}
 Although \eqref{eq:mf1} is only a (1+1)-dimensional system, the presence of the soliton angle $q$ ensures that the resulting modulation solutions, upon reconstructing the corresponding solutions of the KP equation \eqref{eq:kp}, have a non-trivial two-dimensional structure.
 \subsection{Properties of the modulation equations}
 \label{sec:prop}
For our purposes, the most important mathematical feature of the modulation system \eqref{eq:mf1} is its diagonalizability. The system has three Riemann invariants, quantities that are constant along characteristic curves, and thus are coordinates in which the modulation equations \eqref{eq:mf1} are diagonal. The diagonalizability of a 3-component system is quite nontrivial \cite{whitham}. In our case, this diagonalization is consistent with the complete integrability of the KP equation \cite{biondini2019integrability}. An examination of the eigenvalues of \eqref{eq:mf1} reveals that the system is strictly hyperbolic apart from three planes, two of which represent reductions in the number of modulation equations, further evidence of the equation's special structure \cite{dafermos_hyperbolic}.  
\par The eigenvalues of the coefficient matrix in \eqref{eq:mf1} are 
\begin{equation}
  \label{eq:mf2}
  \lambda_{\ub} = \ub, \qquad \lambda_+ = \ub + \frac{a}{3} -  q^2 -
  \frac{2}{3} q\sqrt{ a }, \qquad \lambda_- = \ub + \frac{a}{3} -  q^2 +
  \frac{2}{3} q\sqrt{a} .
\end{equation}
The corresponding Riemann invariants and diagonal form of \eqref{eq:mf1} are 
\begin{align}
    \label{eq:mf3}
    R_{\ub} &= \ub, \qquad  R_\pm = \ub + \frac{1}{2}(q\pm \sqrt{a})^2, \qquad     \frac{\partial R_{j}}{\partial t}+\lambda_{j}\frac{\partial R_{j}}{\partial x}=0, \quad j \in \{\ub,+,-\} \\
    \label{eq:mf2b}
      \lambda_{\ub} &= R_{\ub}, \qquad \lambda_{\pm} = \frac{5}{3}R_{\ub}-\frac{2}{3}\left(R_{\pm}-2\sigma \sqrt{(R_+-R_{\ub})(R_- - R_{\ub})} \right), 
\end{align}
where $\sigma=\sgn(a-q^2)$. The eigenvalues \eqref{eq:mf2} are always real and distinct outside of $q^2 \in \{ 0,\frac{1}{9}a,a \}$, where two eigenvalues coalesce, so the system is hyperbolic everywhere but only strictly hyperbolic outside this set \cite{biondini2019integrability}. At $q^2 \in \{ 0,a \}$ reduced cases exist, since two Riemann invariants coalesce along with the eigenvalues. Note also that the mean flow in \eqref{eq:mf1} is entirely decoupled and  is itself a Riemann invariant in \eqref{eq:mf3}.
\par Due to the symmetry of the KP equation \eqref{eq:kp} and modulation system \eqref{eq:mf1}, throughout this report we will assume that $q>0$ for all initial conditions. Under this assumption, $q$ will remain positive except when strict hyperbolicity is lost at $q=0$, which will be examined below in Section~\ref{sec:ub_zero}. To solve the corresponding problem for $q<0$ we can take $y \rightarrow -y$, which will also lead to $R_\pm \rightarrow R_\mp$.

\subsection{Initial conditions}
\label{sec:ic}
In sections~\ref{sec:riem}--\ref{sec:spec} we 
will study the interaction of a line soliton with a mean flow by looking for solutions to \eqref{eq:mf1} produced by Riemann problems, that is, step-like initial conditions in the modulation variables:
\begin{equation}
    \label{eq:mf_init}
    \ub(x,0)=\begin{cases} \ub_{\rm L} & x<0 \\ \ub_{\rm R} & x>0 \end{cases}, \qquad  
    a(x,0)=\begin{cases} a_{\rm L} & x<0 \\ a_{\rm R} & x>0 \end{cases}, \qquad
    q(x,0)=\begin{cases} q_{\rm L} & x<0 \\ q_{\rm R} & x>0 \end{cases}.
\end{equation}
Once a modulation solution for $\ub(x,t)$, $a(x,t)$, and
$q(x,t)$ is obtained, the modulated soliton is reconstructed by projection onto
\eqref{eq:soli} according to
\begin{equation}
  \label{eq:39}
  \begin{split}
    u(x,y,t) &= \ub(x,t) +  a(x,t) \,\mathrm{sech}^2\left (q(x,t)
      \sqrt{\frac{a(x,t)}{12}} \xi \right ), \\
    \xi &= \int_0^x \frac{1}{q(x',t)}\,\mathrm{d}x'  + y - \int_0^t
    \frac{c(0,t')}{q(0,t')}\,\mathrm{d} t',
  \end{split}
  \end{equation}
  where $c(x,t)$ is defined as in \eqref{eq:soli} with modulated variables. We remark that the modulation equation for $q$ in \eqref{eq:mf1} is a result of the compability condition $\xi_{xt}=\xi_{tx}$. Throughout the remainder of the paper, we will reduce the number of free parameters for the problem \eqref{eq:mf_init} by applying scaling and Galilean symmetries to $\ub_{\rm L}$ and $\ub_{\rm R}$ so that $\ub_{\rm L,R} \in \{0,1\}$, with $\ub_{\rm L}=0$ and $\ub_{\rm R}=1$ for the RW case and $\ub_{\rm L}=1$ and $\ub_{\rm R}=0$ for the DSW case.
\begin{figure}
    \centering
    \includegraphics{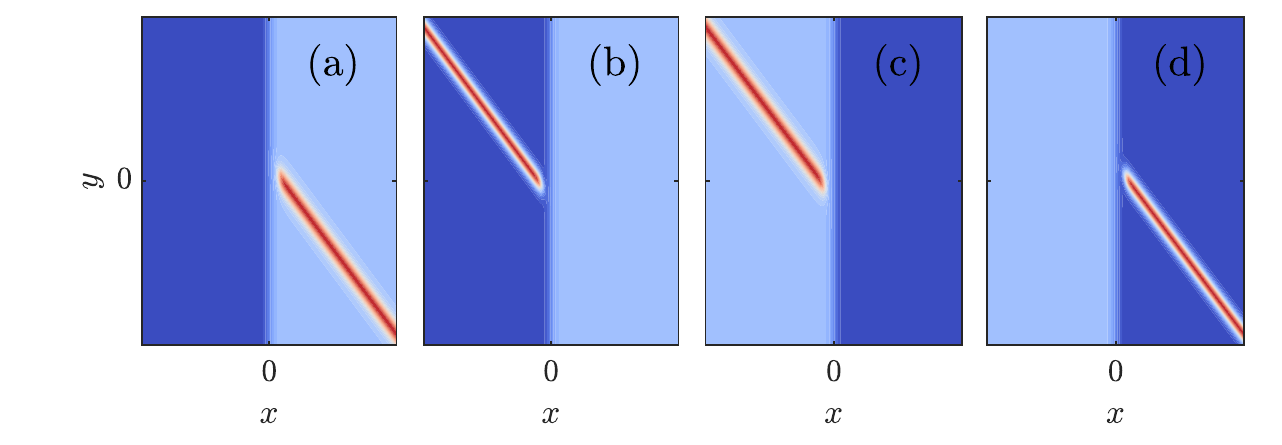}
    \caption{The four initial conditions examined in this report, from left to right: (a) RW-soliton interaction; (b) Soliton-RW interaction; (c) Soliton-DSW interaction; (d) DSW-soliton interaction. Dark blue represents the lower background value.}
    \label{fig:init_cond}
\end{figure}
\par In section~\ref{sec:riem}, we find general simple wave solutions to \eqref{eq:mf_init}, which serve as building blocks for various initial conditions. In sections~\ref{sec:gen} and \ref{sec:spec}, we specifically examine the partial soliton--mean flow initial value problem. For this problem, the parameters on the side of the mean flow farthest from the initial soliton, denoted by $(\ub_1,a_1,q_1)$, are fixed for partial soliton initial conditions as
\begin{equation}
    \label{eq:mf_prt_soli_ic}
    a_1 = 0, \qquad q_1 = q_*,
\end{equation}
where $q_*$ is determined as part of the solution. There are four categories initial value problems to be considered. The initial soliton parameters $(\ub_0,a_0,q_0)$ can be given on either the left or the right. The step in the mean flow can either be upward or downward, where $\ub_{\rm L}<\ub_{\rm R}$ leads to a RW solution, while $\ub_{\rm L}>\ub_{\rm R}$ generates a DSW. Consequently, the four types of interactions to be considered are as follows, where the naming convention differentiates cases where the discontinuity in the mean flow gives rise to a rarefaction wave or a dispersive shock wave
and the component names are ordered 
depending on whether the initial soliton is to the left or to the right of the discontinuity in the mean flow:
\begin{enumerate}
    \item RW-soliton: $\ub_{\rm L}<\ub_{\rm R}$,  $(a_{\rm R},q_{\rm R})=(a_0,q_0)$,  $(a_{\rm L},q_{\rm L})=(0,q_*)$
    \item Soliton-RW: $\ub_{\rm L}<\ub_{\rm R}$,  $(a_{\rm L},q_{\rm L})=(a_0,q_0)$,  $(a_{\rm R},q_{\rm R})=(0,q_*)$
    \item Soliton-DSW: $\ub_{\rm L}>\ub_{\rm R}$,  $(a_{\rm L},q_{\rm L})=(a_0,q_0)$,  $(a_{\rm R},q_{\rm R})=(0,q_*)$
    \item DSW-soliton: $\ub_{\rm L}>\ub_{\rm R}$,  $(a_{\rm R},q_{\rm R})=(a_0,q_0)$,  $(a_{\rm L},q_{\rm L})=(0,q_*)$
\end{enumerate}
These four types of initial conditions are depicted in figure~\ref{fig:init_cond}. Throughout the paper and in all figures, we will refer to the initial conditions by their corresponding letter above (e.g. (a), (b)). These initial conditions present tractable problems that allow for both exact solutions and numerical simulation, enabling a quantitative comparison with our analysis.
\subsection{Reduction to constant mean flow}
\label{sec:const_mean_flow}
For constant $\ub$, the system \eqref{eq:mf1} reduces to 
\begin{equation}
    \label{eq:red_mean_flow}
       \begin{bmatrix}
     a \\ q
  \end{bmatrix}_t +
  \begin{bmatrix}
     \ub+\frac{a}{3} - q^2 & - \frac{4}{3}
     a q 
    \\
    -\frac{1}{3} q & \ub+\frac{a}{3} - q^2
  \end{bmatrix}
  \begin{bmatrix}
    a \\ q
  \end{bmatrix}_x = 0 .
\end{equation}
This reduced system has the Riemann structure:
\begin{equation}
    \label{eq:ub_zero_1}
    r_{\pm}=q \pm \sqrt{a}, \qquad \Lambda_\pm = \ub+\frac{a}{3}-q^2 \mp \frac{2}{3}q\sqrt{a}, \qquad \frac{\partial r_\pm}{\partial t} + \Lambda_\pm \frac{\partial r_\pm}{\partial x} = 0,
\end{equation}
where $r_{\pm}$ are Riemann invariants and $\Lambda_\pm$ are  eigenvalues for the reduced system \eqref{eq:red_mean_flow}. Strict hyperbolicity is lost only at $q=0$. 
The reduced system~\eqref{eq:red_mean_flow} will be used in section~\ref{sec:ub_zero} to study 
the evolution of a truncated soliton, which in turn will provide a stepping stone to study 
the more complicated initial conditions~\eqref{eq:mf_init} subject to \eqref{eq:mf_prt_soli_ic} in section~\ref{sec:spec}.
It will be helpful to recognize that the Riemann invariants in \eqref{eq:ub_zero_1} coincide with 
certain solution parameters in the Wronskian representation of the multi-soliton solutions of the KP equation. We discuss this connection next.
 
\subsection{Multi-soliton solutions of the KP equation, Riemann invariants and soliton interactions}
It was shown in \cite{kodama_book,Biondini_2006} that a large class of multi-soliton solutions of the KPII equation 
can be expressed using the Wronskian representation.
In particular, for \eqref{eq:kp}, one has the solution:
\begin{equation}
u(x,y,t) = 12\, \frac{\partial^2}{\partial x^2}[\log\tau(x,y,t)]\,, 
\label{e:ugeneral}
\end{equation}
where the so-called tau function $\tau(x,y,t)$, 
is given by
\begin{subequations}
\begin{equation}
\tau(x,y,t) = \Wr(f_1,\dots,f_N)\,, 
\end{equation}
the functions $f_1,\dots,f_N$ are given by 
\vspace*{-1ex}
\begin{equation}
f_n(x,y,t) = \sum_{m=1}^M A_{n,m} e^{\theta_m(x,y,t)}\,,
\end{equation}  
and the phases $\theta_1,\dots,\theta_M$ are 
\begin{equation}
\theta_m(x,y,t) =  K_m x + \sqrt3\,K_m^2 y -4K_m^3 t + \theta_{m,0}\,.
\end{equation}
\end{subequations}
The above solution is uniquely determined by the phase parameters $K_1,\dots,K_M$ and the coefficient matrix $A = (a_{m,n})$,
plus the translation constants $\theta_{1,0},\dots,\theta_{M,0}$.
Without loss of generality, one can take the phase parameters to be ordered so that $K_1<\cdots<K_M$.

It was shown in \cite{Biondini_2006} that, generically,
the above representation produces a solution with exactly $N$ asymptotic line solitons as $y\to\infty$
and $M-N$ asymptotic line solitons as $y\to-\infty$.
Such solutions are labeled $(M{-}N,N)$-soliton solutions.
The amplitude and slope of each asymptotic line soliton are completely determined by the
phase parameters $K_1,\dots,K_M$, but the precise details depend on the specific choice of the coefficient matrix $A$.
In the simplest nontrivial case, obtained when $N =1$ and $M=2$, one recovers the soliton solution~\eqref{eq:soli} with $\bar u=0$.
It is convenient to label the two phase parameters as $K_-$ and $K_+$ in this case.
The amplitude and slope parameters $a$ and $q$ are given by 
\begin{equation}
a = 3 (K_+ - K_-)^2\,,\qquad
q = \sqrt3\,(K_+ + K_-)\,.
\label{e:directmap}
\end{equation}
The inverse map to \eqref{e:directmap} is
\begin{equation}
K_- = ( q - \sqrt{a})/(2\sqrt3)\,\qquad
K_+ = ( q + \sqrt{a})/(2\sqrt3)\,.
\label{e:inversemap}
\end{equation}
Comparing~\eqref{e:inversemap} with the first of~\eqref{eq:ub_zero_1} we see that, 
apart from a trivial rescaling, the phase parameters in the Wronskian representation of the multi-soliton solutions
of the KPII equation are precisely the Riemann invariants of the constant mean soliton modulation system~\eqref{eq:red_mean_flow}.

The fact
that the phase parameters in the multi-soliton solutions coincide with the Riemann invariants of the soliton modulation system 
has important ramifications for this work.  
Suppose that one wants to construct a multi-soliton solution consisting of two line solitons with 
amplitude and slope parameters $(a_1,q_1)$ and $(a_2,q_2)$.  
The two sets of phase parameters, one set associated to each soliton, are, respectively, $K_{1,\pm}$ and $K_{2,\pm}$
as given by~\eqref{e:inversemap}.
Importantly, it was shown in~\cite{Kodama_2004} that the resulting two-soliton solution differs
depending on the relative ordering of $K_{2,\pm}$ compared to $K_{1,\pm}$. Further, it was also shown in~\cite{Biondini_2007}
that each of the above three cases corresponds to a different kind of soliton interaction. Specifically, taking $K_{1,-}<K_{2,-}$ without loss of generality, there are three different classes of solutions,
corresponding to the following three possible cases and interactions:
\begin{enumerate}[label=(\roman*)]
    \item Ordinary soliton interaction:  $K_{1,+} < K_{2,-}$ if and only if $\sqrt{a_1}+\sqrt{a_2} < |q_1 - q_2|$,
    \item Resonant soliton interaction: $K_{2,-} < K_{1,+} < K_{2,+}$ if and only if $ |\sqrt{a_1}-\sqrt{a_2}| < |q_1 - q_2| < \sqrt{a_1}+\sqrt{a_2}$, and
    \item Asymmetric soliton interaction: $K_{2,+} < K_{1,+}$ if and only if $|q_1 - q_2| < |\sqrt{a_1}-\sqrt{a_2}|$.
\end{enumerate}

Importantly, in section~\ref{sec:spec}, we will show that each of the above three kinds of soliton interactions arise as a result of 
the time evolution of a particular subset of the class of initial conditions discussed in section~\ref{sec:ic}.

\section{Simple wave solutions}
\label{sec:riem}
In this section, we utilize the mathematical structure of the modulation system \eqref{eq:mf1} outlined in Section~\ref{sec:prop} and \ref{sec:const_mean_flow} to study simple wave solutions for special cases of the initial value problem \eqref{eq:mf_init}. First, we examine the cases $a\equiv 0$ and $q \equiv 0$, both of which represent problems that can be formulated using the KdV equation and have been previously solved. Next, we consider a partial soliton with a constant mean flow $\ub \equiv const.$ by solving the corresponding Riemann problem explicitly. Under certain conditions, the solution is undefined (multivalued), which we resolve by solving a Riemann problem for the (2+1)-dimensional modulation equations. Finally, we look for simple wave solutions to \eqref{eq:mf_init} where $\ub_L \neq \ub_R$. These simple wave solutions are building blocks for solving more complex initial value problems, and they will be utilized as such later in this paper.

\subsection{KdV reductions}
\label{sec:kdv_red}
When $a \equiv 0$ for the initial conditions \eqref{eq:mf_init}, we have simply a KdV Riemann problem in $x$, since $q$ for zero amplitude is undefined. When $\ub_{\rm L}=0$ and $\ub_{\rm R}=1$, a centred RW arises that is defined by 
\begin{equation}
    \label{eq:kdv_rw}
    \ub(x,t) = \begin{cases} 
    1 & t < x\\
    x/t &  0<x<t \\
    0 & x< 0
            \end{cases}.
\end{equation}  The case with $\ub_{\rm L}>\ub_{\rm R}$ is significantly more complex. The initial conditions are compressive, and the corresponding singularity is regularized by dispersion, resulting in a DSW \cite{dsw_review_2016}. The modulation solution is known as the Gurevich-Pitaevski solution \cite{gurevich_1974}.  A DSW consists of a rank-ordered oscillatory train, where for $\ub_{\rm L}=1$ and $\ub_{\rm R}=0$, the leading edge is approximately a  soliton with amplitude $a=2$ (and $q=0$ for KP) and the trailing edge consists of modulated, vanishing harmonic waves. The DSW rightmost leading edge has velocity $\frac{2}{3}$, and the trailing edge has velocity $-1$.
\begin{figure}
    \centering
    \includegraphics[scale=.25,trim={0 3cm 0 0},clip]{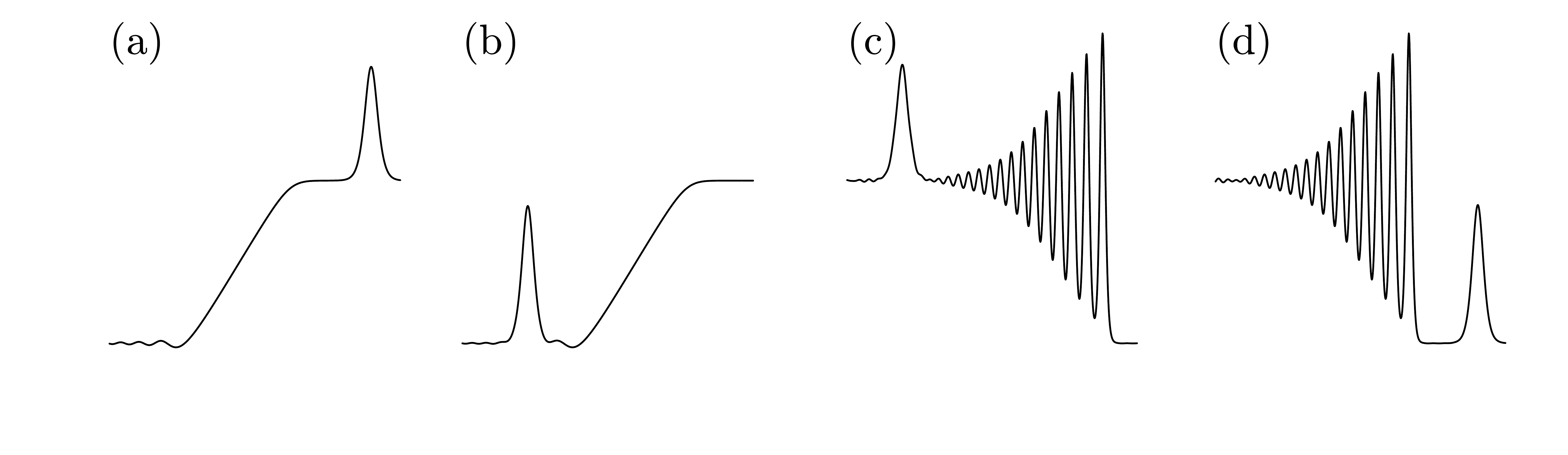}
    \caption{Four types of KdV soliton--mean flow interactions. These are one-dimensional analogues to the initial conditions shown in figure~\ref{fig:init_cond}. Panels (a) and (b) show a soliton interacting with a RW from the right and left, respectively. Panels (c) and (d) show a soliton interacting with a DSW from the left and right, respectively. }
    \label{fig:one_dim}
\end{figure}
\par When $q \equiv 0$ while $a$ and $\ub$ vary, the KP soliton reduces to the KdV soliton, and the reduced (1+1)-dimensional problem can be entirely described using the KdV equation. The KdV soliton--mean flow interaction problem has been studied previously, both using the inverse scattering transform \cite{ablowitz_cole_2018} and Whitham modulation theory \cite{maiden_hoefer_2018}. There are four types of configurations, shown in figure~\ref{fig:one_dim}, which are one-dimensional analogues of the four initial conditions listed in section~\ref{sec:ic} and depicted in figure~\ref{fig:init_cond}.  The findings for KdV are as follows:
\begin{enumerate}
    \item RW-Soliton: soliton does not interact with the RW
    \item Soliton-RW: soliton transmits through the RW if $a_0>2$, otherwise it is trapped
    \item Soliton-DSW: soliton always transmits through DSW
    \item DSW-Soliton: soliton does not interact with the DSW if $a_0>2$, otherwise it is trapped
\end{enumerate}
\par One important result for one-dimensional soliton--mean flow interactions is the principle of \emph{hydrodynamic reciprocity}. Adiabatically invariant quantities are conserved globally, except within the interior of a DSW. Thus, a soliton transmitting through a box, an upward step followed by a downward step of equal magnitude, will return to its original amplitude, subject to a phase (position) shift. This is a consequence of the time-reversability of the governing equation and the continuity of solutions to the Whitham modulation equations. Initial conditions leading to the development of a DSW for $t>0$ will lead to a RW for $t<0$; thus, the same analysis can be applied to both soliton--mean evolutions outside the DSW.

\subsection{Constant mean flow}
\label{sec:ub_zero}
In this section, we consider the evolution of a partial soliton with a constant mean flow $\ub \equiv const.$ This problem has been previously solved using the $x$-independent modulation equations in \cite{ryskamp_2020,neu_singular_2015}. We will solve this problem using \eqref{eq:red_mean_flow}, the constant mean reduction of the $y$-independent modulation equations \eqref{eq:mf1}. This solution will prove to be a building block for solutions of the full Riemann problem \eqref{eq:mf_init} subject to \eqref{eq:mf_prt_soli_ic}. It will also reveal the necessity and utility of a multivalued solution in $x$, a novel feature of the problem. In \ref{sec:eq_simp_waves} we show that solving the $x$- and $y$-independent modulation equations yields equivalent solutions for the partial soliton. There are two cases to consider, depending on whether the partial soliton starts to the left or to the right.

\paragraph{Partial soliton on the right} 
Let us first consider the Riemann problem with the partial soliton on the right:
\begin{equation}
    \label{eq:ub_zero_2}
       a_{\rm p}(x,0)=\begin{cases} 0 & x<0 \\ a_0 & x>0 \end{cases}, \qquad
    q_{\rm p}(x,0)=\begin{cases} q_* & x<0 \\ q_0 & x>0 \end{cases},
\end{equation}
where $q_0>0$. As the soliton modulation is expanding into the $a=0$ ``vacuum", the resulting solution is sought in the form of a simple wave. The vacuum region is to the left of the nonzero region, and since $\Lambda_+<\Lambda_-$, we seek a $r_-$-wave in which $r_+$ is constant. Thus, $r_+$ determines $q_*$ as
\begin{equation}
    \label{eq:qstar1}
    r_+=q_*=q_0+\sqrt{a_0},
\end{equation} while the simple wave that develops is a 2-wave with $r_-$ changing.  We can solve for the simple wave solution
\begin{equation}
\begin{split}
  \label{eq:ub_const_sw}
    q_{\rm p}(x,t) &=
    \begin{cases}
      q_0 & \quad U_{\rm s} t < x \\  \frac{1}{2}\left[q_*^2+3(\ub-\frac{x}{t})\right]^{1/2} & \enskip \;
      U_{\rm z} t < x < U_{\rm s} t \\ q_* & \qquad\quad\, \, x < U_{\rm z} t
    \end{cases}, \\ \sqrt{a_{\rm p}(x,t)} &= q_0+\sqrt{a_0} - q_{\rm p}(x,t) ,\end{split}
\end{equation}
where the characteristic velocities are 
\begin{equation}
    \label{eq:ub_const_sw_2}
    U_{\rm s} = \ub+\frac{a_0}{3}-q_0^2+\frac{2}{3}q_0\sqrt{a_0}, \qquad U_{\rm z} = \ub-q_*^2.
\end{equation}
We call $U_{\rm s}$ the velocity of the \emph{soliton edge} of the simple wave and $U_{\rm z}$ the velocity of the \emph{zero edge} of the simple wave. We will be referring to these edges and velocities often. Note that the zero edge of the partial soliton simple wave always moves left with respect to the mean flow, while the soliton edge may move left or right relative to the mean flow, depending on the parameter values.
\begin{figure}
    \centering
    \includegraphics{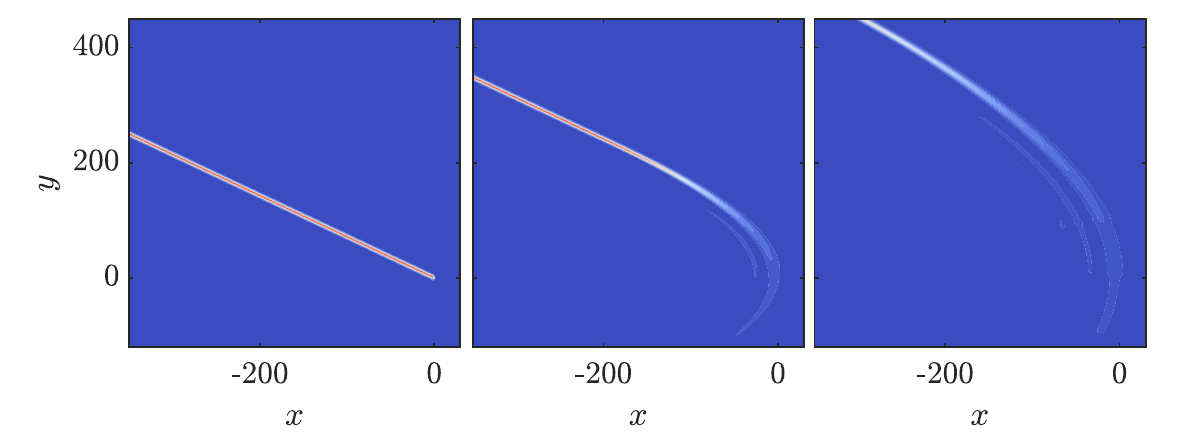}
    \caption{Numerical evolution of the KP equation \eqref{eq:kp} for a partial soliton with initial conditions \eqref{eq:ub_zero_2} with $\ub=0$, $\sqrt{a_0}=\sqrt{a_{\rm L}}=1<q_0=q_{\rm L}=1.4$ for $t \in (0,60,150)$. For large $t$, the one-dimensional modulation solution is well-defined, i.e. it is not multivalued in $x$.}
    \label{fig:partial_evolution_not_mult}
\end{figure}

\paragraph{Partial soliton on the left}
Let us now consider the reflected Riemann problem, where the soliton starts on the left,
\begin{equation}
    \label{eq:ub_zero_3}
       a_{\rm p}(x,0)=\begin{cases} a_0 & x<0 \\ 0 & x>0 \end{cases}, \qquad
    q_{\rm p}(x,0)=\begin{cases} q_0 & x<0 \\ q_* & x>0 \end{cases},
\end{equation} and again $q_0>0$. In this case, the vacuum state is to the right, so $r_-$ is constant, determining $q_*$ as
\begin{equation}
\label{eq:qstar2}
r_-=q_*=q_0-\sqrt{a_0}.
\end{equation} The simple wave that then develops is a $r_+$-wave (or 1-wave) with $r_+$ changing. From \eqref{eq:qstar2}, we have two cases that depend on the sign of $q_*$. If $q_0>\sqrt{a_0}$, then $q_*>0$ and the solution for $q_{\rm p}(x,t)$ has the form
\begin{equation}
  \label{eq:ub_const_sw_large_q}
  \begin{split}
    q_{\rm p}(x,t) &=
    \begin{cases}
      q_* & \quad U_{\rm z} t < x \\  \frac{1}{2}\left[q_*^2+3(\ub-\frac{x}{t})\right]^{1/2} & \enskip \;
      U_{\rm s} t < x < U_{\rm z} t \\ q_0 & \qquad\quad\, \, x < U_{\rm s} t
    \end{cases}, \\
    \sqrt{a_{\rm p}(x,t)} &= -q_0+\sqrt{a_0} + q_{\rm p}(x,t) ,\end{split}
\end{equation}with soliton and zero edge characteristic velocities, respectively, \begin{equation}
    \label{eq:ub_const_sw_3}
    U_{\rm s} = \ub+\frac{a_0}{3}-q_0^2-\frac{2}{3}q_0\sqrt{a_0}, \qquad U_{\rm z} = \ub-q_*^2.
\end{equation}
 A simulation of this case is shown in figure~\ref{fig:partial_evolution_not_mult}.
 \begin{figure}
    \centering
    \includegraphics{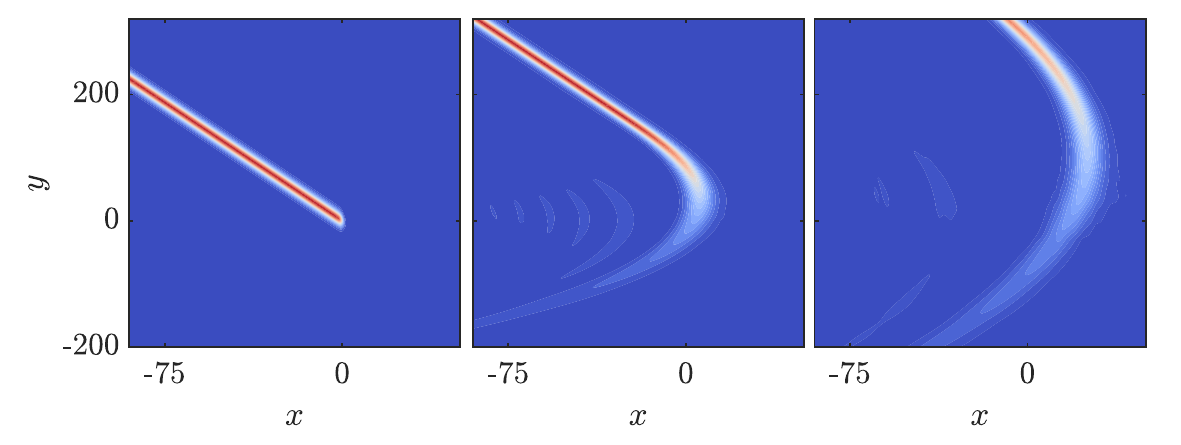}
    \caption{Numerical evolution of the KP equation \eqref{eq:kp} for a partial soliton with initial conditions \eqref{eq:ub_zero_3} with $\ub=0$, $\sqrt{a_0}=\sqrt{a_{\rm L}}=1>q_0=q_{\rm L}=.4$ on $t \in (0,80,250)$. The modulation solution becomes multivalued in $x$.}
    \label{fig:partial_evolution}
\end{figure}
\par If $q_0<\sqrt{a_0}$, then $q_*<0$ and the solution is more complicated. In order to have a continuous solution for $q$, then there must exist some $x$ such that $q_{\rm p}(x,t)=0$, where the system loses strict hyperbolicity. From numerical simulation, presented in figure \ref{fig:partial_evolution}, it is evident the modulation solution becomes multivalued in $x$, and it does so at the branch point $q=0$. 
\par However, we can still construct a well-defined  solution if we appeal to its two-dimensional structure. The simple wave solution for $q$ is inherently multivalued
\begin{equation}
    \label{eq:ub_const_multval}
    q_{\rm p}^{\pm}(x,t) = \pm \frac{1}{2}\left[q_*^2+3\left(\ub-\frac{x}{t}\right)\right]^{1/2},
\end{equation}
which suggests that the true solution can be pieced together by carefully choosing the correct sign for \eqref{eq:ub_const_multval} as a function of $y$. The rightmost part of the wave is where $q(x,t)=0$. As shown in \eqref{eq:zero_speed_in_y} of \ref{sec:eq_simp_waves}, the location where $q=0$ is moving with a velocity in $y$ of $V_{\rm f} = -2 r_-/3 = -2q_* /3$ and, by inspection of \eqref{eq:ub_const_multval}, a velocity in $x$ of $U_{\rm f} = \ub + q_*^2/3$. Thus, the full solution becomes a combination of the two branches. For $y>V_{\rm f} t$,
\begin{subequations}
\label{eq:ub_const_sw_f}
\begin{equation}
  \label{eq:ub_const_sw_fa}
    q_{\rm p}^+(x,t) =
    \begin{cases}
      q_0 & \qquad \, \; x<U_{\rm s} t \\  \frac{1}{2}\left[q_*^2+3\left(\ub-\frac{x}{t}\right)\right]^{1/2} &
      U_{\rm s} t < x < U_{\rm f} t
    \end{cases},
\end{equation}
while for $y<V_{\rm f}t$,
\begin{equation}
  \label{eq:ub_const_sw_fb}
    q_{\rm p}^-(x,t) =
    \begin{cases}
      -\frac{1}{2}\left[q_*^2+3\left(\ub-\frac{x}{t}\right)\right]^{1/2} & \quad U_{\rm z} t < x < U_{\rm f} t \\
      q_* = q_0-\sqrt{a_0} & \qquad\quad\; \; x < U_{\rm z} t
    \end{cases},
\end{equation}\end{subequations}
with the equation for the amplitude the same as \eqref{eq:ub_const_sw_large_q} supplemented with $a=0$ for $x>U_{\rm z}t$ and characteristic velocities the same as \eqref{eq:ub_const_sw_3}.
\par
\begin{figure}
    \centering
    \includegraphics[scale=.25]{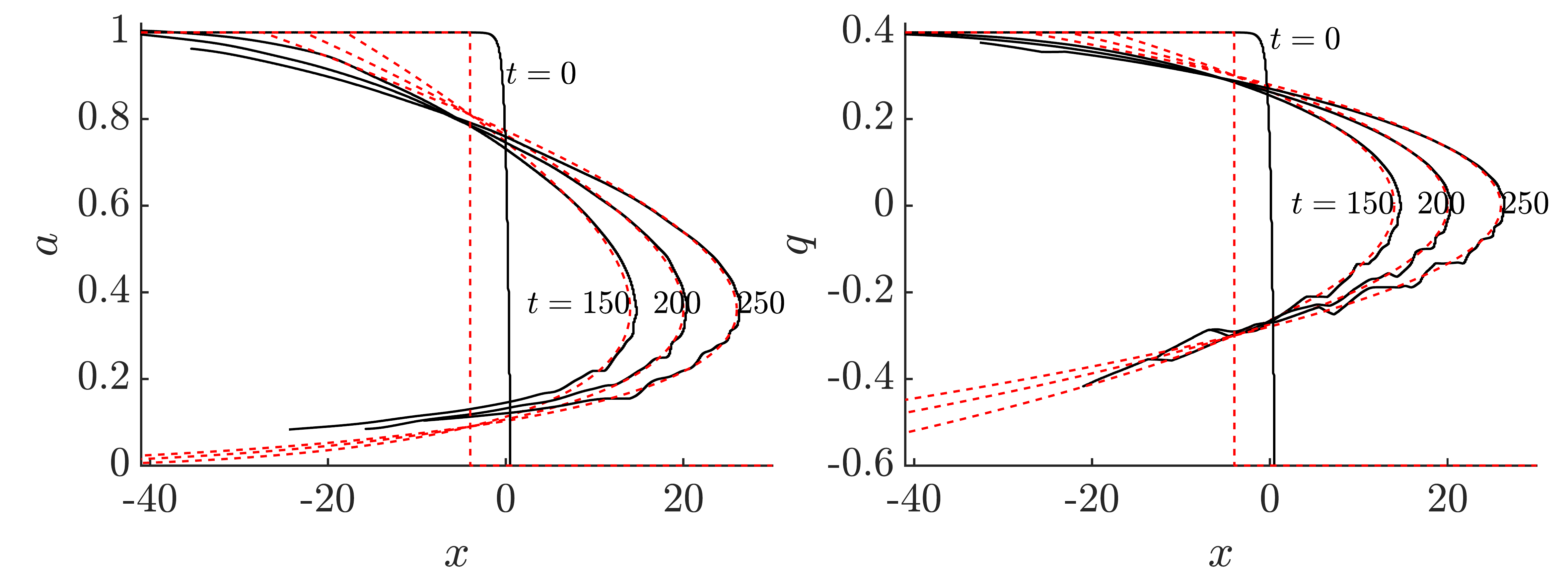}
    \caption{Comparison of analytical solution (dashed) in \eqref{eq:ub_const_sw_f} with numerical simulation (solid) for the same simulation as figure~\ref{fig:partial_evolution} when $t \in (0,150,200,250)$. The analytical solution is shifted by $x_0=-4$.}
    \label{fig:partial_soli_amp_slope}
\end{figure}

 We justify the (2+1)-dimensional modulation solution using two approaches. First, direct numerical analysis confirms our analytical prediction. Figure~\ref{fig:partial_soli_amp_slope} shows good agreement between the simulation in figure~\ref{fig:partial_evolution} and the analytical result \eqref{eq:ub_const_sw_f}. Second, the partial soliton Riemann problem \eqref{eq:ub_zero_3} can be rewritten as the previously solved Riemann problem for the parameters $a$ and $q$ using $x$-independent modulation equations. In that case there is no loss of strict hyperbolicity nor a multivalued evolution. In \ref{sec:eq_simp_waves}, we show that the above solution \eqref{eq:ub_const_sw_f} is equivalent to the modulation solution using the single-valued $x$-independent modulation solution.

\subsection{Soliton--mean simple wave}
\label{sec:sw_soln}
We now look for simple wave solutions to the full Riemann problem \eqref{eq:mf_init}, where parameters are chosen so that $R_{\pm}$ are held constant while $R_{\ub}$ is changing. We will call these solutions \emph{soliton--mean simple waves}. They represent admissible, stable solutions that are expected to be large $t$ attractors for a variety of initial conditions. The parameters $(\ub_0,a_0,q_0)$ are given on the initial soliton side of the discontinuity (left or right) and connect to $(\ub_1,a_1,q_1)$ on the far side (right or left, respectively). For convenience, let $\Delta = \ub_0-\ub_1=\pm 1$. Also, throughout the remainder of this paper, instead of utilizing the conventional numbering of simple waves based on the ordering of characteristic velociies (e.g. 1-wave or 2-wave), we will refer to a simple wave where only $R_j$ is changing as an $R_j$-simple wave, where $R_i$, $i \neq j$, are constant for $i,j \in \{\ub,+,-\}$.
\par An $R_{\ub}$-simple wave gives rise to the two relationships between the parameters for the far and mean sides. Given constant $R_{\pm}$,
\begin{subequations}
    \label{eq:mf6}
\begin{align}
    \label{eq:mf6a}
    R_- = \ub_0+\frac{1}{2}(q_0-\sqrt{a_0})^2 &= \ub_1+\frac{1}{2}(q_1-\sqrt{a_1})^2, \\
     \label{eq:mf6b}
    R_+ = \ub_0+\frac{1}{2}(q_0+\sqrt{a_0})^2 &= \ub_1+\frac{1}{2}(q_1+\sqrt{a_1})^2.
\end{align}
\end{subequations}
Subtracting \eqref{eq:mf6a} from \eqref{eq:mf6b} and squaring yields
\begin{equation}
    \label{eq:mf7}
    \left(\frac{R_+-R_-}{2}\right)^2=q_0^2 a_0=q_1^2 a_1.
\end{equation}
Adding \eqref{eq:mf6a} and \eqref{eq:mf6b} together gives the relation
\begin{equation}
    \label{eq:mf7_2}
    2\Delta+q_0^2+a_0=q_1^2+a_1.
\end{equation}
Eliminating $q_1^2$ in \eqref{eq:mf7} with \eqref{eq:mf7_2}, we obtain two solutions of \eqref{eq:mf6} for $a_1$ 
\begin{subequations}
\label{eq:mf8}
\begin{equation}
    \label{eq:mf8a}
    a_1=\frac{1}{2} \left( a_0+ q_0^2+2\Delta + \sigma \sqrt{(a_0+q_0^2+2\Delta)^2-4 a_0 q_0^2} \right),
\end{equation} 
where $\sigma=\pm 1$. Due to the symmetry of the equations, repeating the above process for $q_1^2$ gives the same result, but with opposite sign
\begin{equation}
    \label{eq:mf8b}
    q_1^{2}=\frac{1}{2} \left( a_0+ q_0^2+2\Delta - \sigma \sqrt{(a_0+q_0^2+2\Delta)^2-4 a_0 q_0^2} \right).
\end{equation} \end{subequations} 
The opposite signs in front of $\sigma$ in \eqref{eq:mf8} are required to satisfy \eqref{eq:mf7_2}. We will define $\sigma$ below. A corollary of \eqref{eq:mf8} is that
\begin{equation}
    \label{eq:10}
    a_1-q_1^{2}=2\sigma\sqrt{(a_0+q_0^2+2\Delta)^2-4 a_0 q_0^2},
\end{equation}
which implies that if $\sigma=1$, $a_1\geq q_1^2$, while if $\sigma=-1$, $a_1\leq q_1^2$. Equality only occurs when $a_0=(\sqrt{-2 \Delta} \pm q_0)^2$, which requires $\Delta<0$. 
\par To have a real expression for \eqref{eq:mf8}, the term underneath the square root must be nonnegative. This happens when one of the following holds
\begin{subequations}
   \label{eq:nonneg}
   \begin{align}
    \label{eq:nonneg_a}
    -2 \Delta &> (q_0+\sqrt{a_0})^2,\\
    \label{eq:nonneg_b}
    -2 \Delta &< (q_0-\sqrt{a_0})^2.
\end{align}
\end{subequations}
In addition, for soliton transmission (defined below) to be well-defined, both $a_1$ \eqref{eq:mf8a} and $q_1^2$ \eqref{eq:mf8b} must be positive. This occurs when
\begin{equation}
    \label{eq:trans_cond}
    -2\Delta < (q_0-\sqrt{a_0})^2,
\end{equation}
an identical requirement to \eqref{eq:nonneg_b}. Thus, \eqref{eq:trans_cond} completely describes the transmission regime for a KP soliton and is a necessary and sufficient condition to guarantee a simple wave solution across a changing mean flow. Note that setting $q_0=0$ and $\Delta=-1$ reduces \eqref{eq:trans_cond} to the transmission condition described in section~\ref{sec:kdv_red} for both a soliton-RW interaction and a DSW-soliton, namely that a soliton is trapped when $a_0<2$. This motivates the following:
\begin{definition}
\label{def:trapping}
A soliton interacting with a changing mean flow is \emph{trapped} if \eqref{eq:trans_cond} does not hold, i.e., if $-2 \Delta > (q_0-\sqrt{a_0})^2$.
\end{definition}
We define transmission to be the opposite of trapping:
\begin{definition}
\label{def:transmitting}
A soliton interacting with a changing mean flow is \emph{transmitted} if \eqref{eq:trans_cond} holds, i.e., if $-2 \Delta < (q_0-\sqrt{a_0})^2$.
\end{definition}
\par For a transmitted soliton, we can use conservation of $R_\pm$ to calculate exact solutions for $a(x,t)$ and $q(x,t)$ within a RW. This will also define the sign $\sigma$. Since the exact solution for $\ub(x,t)$ in a RW is known \eqref{eq:kdv_rw} and $R_\pm$ are constant, we can find $a(x,t)$ and $q(x,t)$ in a soliton--mean simple wave with a RW by solving the system of equations
\begin{equation*}
    \label{eq:mf19}
     R_{+} = \frac{x}{t}+ \frac{1}{2} \left( q(x,t)+\sqrt{a(x,t)} \right)^2, \qquad R_{-} = \frac{x}{t}+ \frac{1}{2}\left( q(x,t)-\sqrt{a(x,t)} \right)^2,
\end{equation*}
where $R_{\pm}$ are the constant values $R_{\pm}=\ub_0+\frac{1}{2}(q_0\pm\sqrt{a_0})^2=\ub_1+\frac{1}{2}(q_1\pm\sqrt{a_1})^2$. Solving these equations in a similar manner as above gives explicit solutions for $a(x,t)$ and $q(x,t)$,
\begin{subequations}
\label{eq:mf11}
\begin{align}
    \label{eq:mf11a}
    a(x,t) &=-\frac{x}{t}+\frac{R_{+} + R_{-}}{2} + \sigma \left[\left( \frac{x}{t}-R_{-} \right) \left( \frac{x}{t}-R_{+} \right)\right]^{1/2}, \\
    \label{eq:mf11b}
    q^{2}(x,t) &=-\frac{x}{t}+\frac{R_{+} + R_{-}}{2} - \sigma \left[\left( \frac{x}{t}-R_{-} \right) \left( \frac{x}{t}-R_{+} \right)\right]^{1/2}.
    \end{align}
\end{subequations} 
At the edge of the RW closest to the initial soliton with parameters $(\ub_0,a_0,q_0)$, from \eqref{eq:kdv_rw} we have that $x/t=\ub_0$, $a(x,\ub_0 t)=a_0$, and $q^2(x,\ub_0 t)=q_0^2$. Inserting these values and subtracting \eqref{eq:mf11b} from \eqref{eq:mf11a} yields, in physical variables,
\begin{equation}
    \label{eq:mf13}
    a_0-q_0^2 = \sigma |a_0-q_0^2|.
\end{equation}
The above relation \eqref{eq:mf13} implies that
\begin{equation}
    \label{eq:sigma}
    \sigma = \sgn(a_0-q_0^2), \quad a_0 \neq q_0^2.
\end{equation}
Thus, $\sigma$ is determined by consistency with the soliton initial conditions $(a_0,q_0^2)$. The determination of $\sigma$ \eqref{eq:sigma} combined with \eqref{eq:10} shows that $a_0 > q_0^2$ implies $a_1 \geq q_1^2$, while $a_0 < q_0^2$ implies $a_1 \leq q_1^2$. In general, if $a>q^2$ anywhere within the soliton--mean simple wave, $a \geq q^2$ throughout the entire solution, and vice versa. 
 \par If $a_0=q_0^2$, the mapping \eqref{eq:mf8} is undetermined in general. That special case is only relevant if $\Delta>0$, since if $\Delta<0$ then $a_0=q_0^2$ will not transmit according to \eqref{eq:trans_cond}. For $\Delta=1>0$, \eqref{eq:10} implies that 
 \begin{equation}
     \label{eq:11}
    \lim_{a_0 \to q_0^{2\pm}} (a_1-q_1^2)=\pm 4 \sqrt{2a_0+1},
 \end{equation}
 which has different values from the $a_0>q_0^2$ ($\sigma=+1$) and $a_0<q_0^2$ ($\sigma = -1$) sides. This discontinuous dependence on initial data implies that the mapping \eqref{eq:mf8} is undetermined for $a_0=q_0^2$ and $\Delta=1>0$, which is unsurprising given the lack of strict hyperbolicity for those parameters. In the remainder of this section, we assume that $a_0 \neq q_0^2$. In section~\ref{sec:sw_mf_int} we examine a specific initial value problem where $a_0=q_0^2$ and give an argument for choosing $\sigma$ in that scenario, a choice validated by numerical simulation.
 
\par Although the above analysis based on \eqref{eq:mf11} assumes a RW, the principle of hydrodynamic reciprocity \cite{maiden_hoefer_2018} referenced in section~\ref{sec:kdv_red} allows us to extend the result \eqref{eq:sigma} to a DSW. If we assume a mean flow resulting in a DSW ($\ub_L>\ub_R$) for $t>0$, the time reversability of the KP equation \eqref{eq:kp} implies that $t<0$ will result in a RW. If we obtain the global solution through the RW for $t<0$ by assuming $R_\pm$ constant as above in \eqref{eq:mf11}, then we have a solution with $R_\pm$ constant \emph{outside the DSW region} also for $t>0$. Consequently \eqref{eq:sigma} also applies to DSW initial conditions.

\par The mapping from \eqref{eq:mf8} across a mean flow is also invertible. By direct evaluation from \eqref{eq:mf8} we calculate that if $\Delta=1$, then
\begin{equation}
    \label{eq:mf15}
    (q_1-\sqrt{a_1})^2 = 2 + (q_0-\sqrt{a_0})^2 > 2.
\end{equation}
This means that for any soliton $(a_0,q_0)$ transmitting from the higher $\Delta=1$ side of a mean flow, its counterpart across the mean flow $(a_1,q_1)$ in \eqref{eq:mf15} with $\Delta=-1$ also satisfies the transmission condition \eqref{eq:trans_cond}. Let us now define the soliton--mean simple wave mapping
\begin{subequations} 
\label{eq:f_map}
\begin{equation}
\label{eq:f_map_1}
\mathcal{F}_{\Delta=1}: \{(a_0,q_0^2) \mid a_0>0, \, q_0>0 \}\rightarrow \{(a_1,q_1^2) \mid a_1>0, \, q_1>0, \, (q_1-\sqrt{a_1})^2>2 \},
\end{equation}
where $\mathcal{F}_{\Delta=1}$ is defined by \eqref{eq:mf8} with $\Delta=1$ and \eqref{eq:sigma}. $\mathcal{F}_{\Delta=1}$ maps transmitted soliton parameters across a mean flow when the initial soliton is posited on the higher ($\Delta=1$) side of the initial step. We also define the mapping
\begin{equation}
\label{eq:f_map_-1}
\mathcal{F}_{\Delta=-1}: \{(a_0,q_0^2) \mid a_0>0, \, q_0>0, \,(q_0-\sqrt{a_0})^2>2 \}\rightarrow \{(a_1,q_1^2) \mid a_1>0, \, q_1>0 \},
\end{equation}\end{subequations}
again defined by \eqref{eq:mf8} and \eqref{eq:sigma} with $\Delta=-1$. $\mathcal{F}_{\Delta=-1}$ maps transmitted soliton parameters across a mean flow when the initial soliton is posited on the lower ($\Delta=-1$) side of the initial step. Direct calculation verifies that  $\mathcal{F}_{\Delta=1}(\mathcal{F}_{\Delta=-1}(a,q^2))=(a,q^2)$. In other words, both $\mathcal{F}_{\Delta=\pm1}$ are one-to-one and onto on their domains, with
 \begin{equation*}
     \mathcal{F}_{\Delta=1}^{-1}=\mathcal{F}_{\Delta=-1}.
 \end{equation*}
\par The soliton--mean simple wave solution described in this section is the main result of this work. The mappings \eqref{eq:f_map} provide a basic condition for existence of a soliton traversing a mean flow. For a soliton to exist across a changing mean flow, the initial parameters must be in the domain of one of \eqref{eq:f_map} and the parameters across the mean flow $(a_1,q_1)$ must be in the range, giving a real and nonnegative result for $a_1$ and $q_1^2$ in \eqref{eq:mf8}. Examples of transmitted solitons fulfilling the mappings \eqref{eq:f_map} are shown in the third panels of figures~\ref{fig:lrw1}, \ref{fig:rrw4}, \ref{fig:rdsw2}, and \ref{fig:ldsw3}. We can also determine transmission or trapping from \eqref{eq:trans_cond}, which generalizes the KdV transmission conditions to the KP equation. In the next two sections we will show how the mappings \eqref{eq:f_map} can be utilized for the specific initial value problem \eqref{eq:mf_init} subject to \eqref{eq:mf_prt_soli_ic}.

\section{General considerations regarding interactions between partial solitons and mean flows}
\label{sec:gen}
We now proceed to discuss the evolution of the partial soliton--mean flow initial data \eqref{eq:mf_init} subject to \eqref{eq:mf_prt_soli_ic} for the modulation equations \eqref{eq:mf1}, whose projection \eqref{eq:39} is shown in figure~\ref{fig:init_cond}. We first consider general properties of the problem, its solution, and criteria for transmission. The transmission conditions here will be more restrictive than those for a single soliton--mean simple wave \eqref{eq:trans_cond}, with additional subtleties due to the multiple waves generated from the partial soliton initial data. Next, we will examine the interaction of a RW with a partial soliton simple wave from section~\ref{sec:ub_zero}, a feature which occurs repeatedly in the analysis of specific cases. In section~\ref{sec:spec} we will apply the following general approach to the four specific initial conditions in figure~\ref{fig:init_cond}.
\subsection{Transmission conditions}
\label{sec:trans_cond}
 We will look for solutions to the Riemann problem \eqref{eq:mf_init} subject to \eqref{eq:mf_prt_soli_ic} consisting of a combination of well-defined simple waves connected by constant states. To facilitate this analysis, let us again refer to the initial soliton parameters, either $(a_{\rm L},q_{\rm L})$ or $(a_{\rm R},q_{\rm R})$, as $(a_0,q_0)$. The mean flow on the side of the initial partial soliton is denoted by $\ub_0$, while the mean flow on the other side of the jump at $x=0$ is $\ub_1$. From \eqref{eq:mf_prt_soli_ic}, we recall our initial data $(a_1,q_1)=(0,q_*)$. Since $q_*$ is chosen to conserve one of $R_\pm$ (cf. \eqref{eq:qstar1} and \eqref{eq:qstar2}), which is also conserved by the soliton--mean simple wave of section~\ref{sec:sw_soln}, transmission is generically characterized by two simple waves connected by one constant state. Since the mean flow $\ub$ is decoupled in \eqref{eq:mf1}, the wave closest to the initial partial soliton will have $R_{\ub}$ changing with both $R_{\pm}$ constant, i.e. the soliton--mean simple wave of section~\ref{sec:sw_soln}. Between the simple waves will be an expanding constant region with parameters $(a_{\rm M},q_{\rm M})$ that are determined by \eqref{eq:f_map}. The other simple wave is for the partial soliton of section~\ref{sec:ub_zero} in which $R_{\ub}$ and only one of $R_{\pm}$ are constant, connecting $(a_{\rm M},q_{\rm M})$ to the constant state $(0,q_*)$. 
  \par Consequently, in order for the partial soliton to completely transmit through a changing mean flow under the initial conditions \eqref{eq:mf_init} and \eqref{eq:mf_prt_soli_ic}, three conditions are necessary. First, the partial soliton must propagate into the mean flow. Second, the partial soliton must transmit through the RW or DSW. Third, the partial soliton on the far side of the changing mean flow must also propagate away from the mean flow. We will consider each of these conditions in turn.

\par First, the partial soliton must propagate into the RW or DSW so that an interaction occurs. The partial soliton simple wave characteristic velocity $U_{\rm s}$ in \eqref{eq:ub_const_sw_2} or \eqref{eq:ub_const_sw_3} must have a magnitude and direction such that the soliton edge of the partial soliton simple wave interacts with the near edge of the mean flow. Specifically, if the partial soliton is initialized to the left of the mean flow, the soliton edge must move faster than the left edge of the RW or DSW, and if the partial soliton is initialized to the right, the right edge of the RW or DSW must overtake the soliton edge. If the partial soliton simple wave soliton edge does not interact with the mean flow we refer to this scenario as either \emph{partial recession} or \emph{total recession}, the former occurring if the zero edge of the partial soliton simple wave interacts with the mean flow.
\begin{definition}
\label{def:recess}
A partial soliton \emph{recedes} from a mean flow if the soliton edge of the partial soliton simple wave defined in section~\ref{sec:ub_zero} never interacts with the mean flow. If the partial soliton simple wave's zero edge also does not interact with the mean flow, this is known as \emph{total recession}; otherwise we call it \emph{partial recession}.
\end{definition}
If $a_0 = a_{\rm R}$ and $U_{\rm s}$ is greater than the right edge velocity of the DSW or RW, since $U_{\rm z}<\ub$ in \eqref{eq:ub_const_sw_2} we will only have partial recession, never total recession. If $a_0=a_{\rm L}$ and $U_{\rm s}$ is less than the left edge velocity of the DSW or RW, since $U_{\rm z}<\ub$ in \eqref{eq:ub_const_sw_3} we will only have total recession. When partial recession occurs, the partial soliton can still transmit or be trapped, while total recession precludes both transmission and trapping. We consider partial recession more fully below in section~\ref{sec:sw_mf_int}. Numerical simulations for predicted conditions showing recession are shown in  figures~\ref{fig:int1} (partial) and \ref{fig:rrw3} (total).
\par Second, the partial soliton interacting with the RW or DSW must then transmit through it. In other words, a well-defined soliton--mean simple wave solution with the appropriate domain and range of the mapping \eqref{eq:f_map} must exist. As defined above in definition~\ref{def:trapping}, when a soliton--mean simple wave solution does not exist, we refer to this as soliton \emph{trapping}. Note that when transmission occurs according to \eqref{eq:trans_cond}, $q_*$ must be well-defined, since 
\begin{equation}
    \label{eq:gen_1}
    q_*^2 = 2\Delta+ (q_0 \pm \sqrt{a_0})^2>0,
\end{equation}
where the sign choice corresponds to whether $R_+$ or $R_-$ is conserved and $\Delta$ is defined as above. Numerical simulations for data predicted to show trapping are shown in figures~\ref{fig:rrw_trap}, \ref{fig:ldsw1}, and \ref{fig:ldsw2}.

\par
\begin{figure}
    \centering
    \includegraphics[scale=.25]{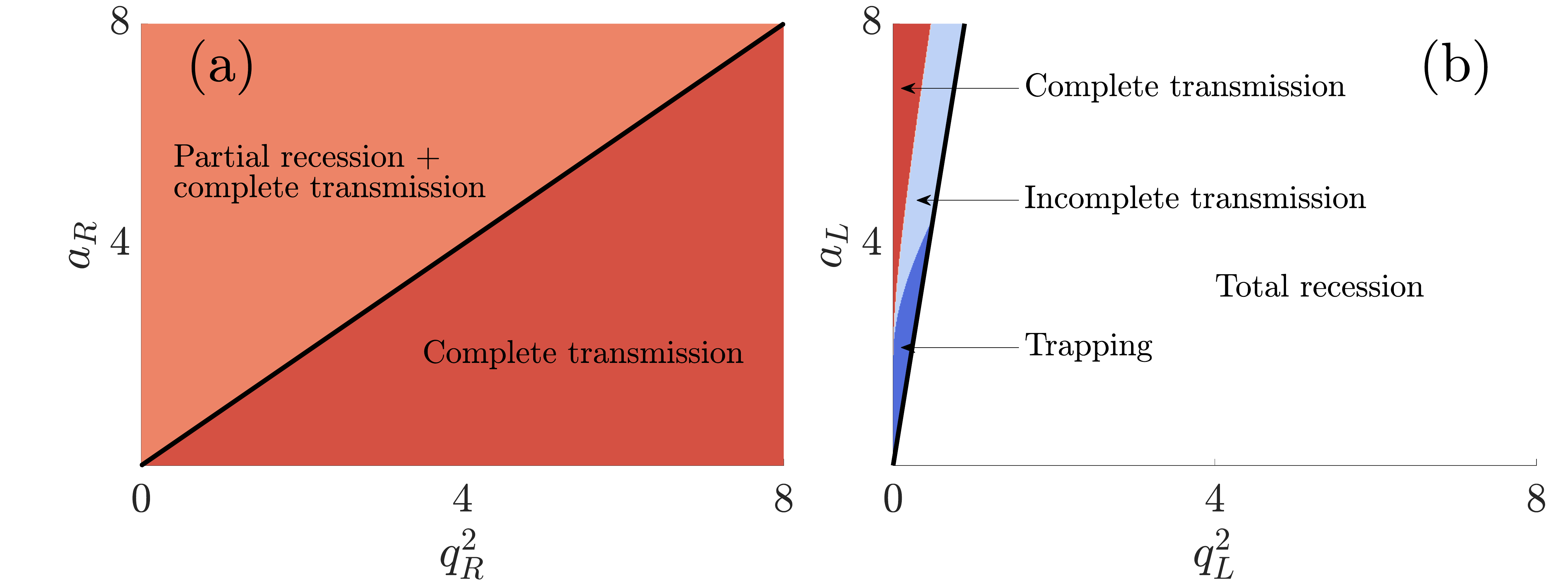}
    \includegraphics[scale=.25]{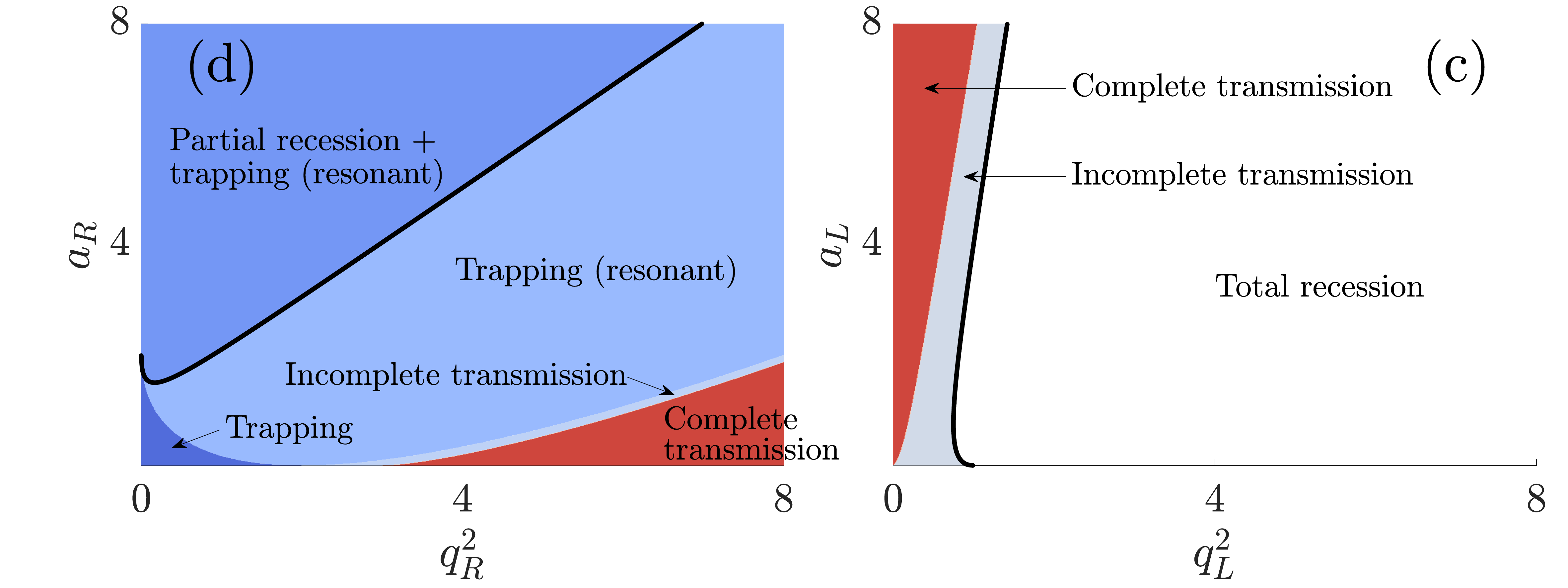}
    \caption{Phase diagram of predicted interactions between a partial soliton with a dynamic mean flow for the four initial conditions shown in figure~\ref{fig:init_cond}: (a) RW--soliton interaction; (b) Soliton--RW interaction; (c) Soliton--DSW interaction; (d) DSW--soliton interaction. For cases (b) and (c), complete transmission also leads to the partial soliton simple wave bending around to interact with the mean flow a second time.}
    \label{fig:transmission region}
\end{figure}

 Third, for the soliton to fully transmit through the RW or DSW, the transmitted partial soliton must continue to propagate away from the mean flow. Assuming the partial soliton transmits through the changing mean flow with $(a_1,q_1)=(a_{\rm M},q_{\rm M})$ in \eqref{eq:mf8}, for the soliton to fully establish itself beyond the DSW or RW there must be a partial soliton simple wave (cf. section~\ref{sec:ub_zero}) connecting $(a_{\rm M},q_{\rm M})$ to $(0,q_*)$ with a sufficiently fast soliton edge velocity $U_{\rm s}$ from \eqref{eq:ub_const_sw_2} or \eqref{eq:ub_const_sw_3}. If the new partial soliton never completely separates from the mean flow, we call this \emph{incomplete transmission}:
\begin{definition}
\label{def:incomplete}
A partial soliton experiences \emph{incomplete transmission} when it transmits through a RW or DSW but the soliton edge of the partial soliton simple wave defined in section~\ref{sec:ub_zero} does not move faster than the nearest edge of the changing mean flow.
\end{definition}
A numerical simulation for data we predict to lead to incomplete transmission is shown in figure~\ref{fig:rrw2}. If a fully established line soliton separates from the mean flow on the far side, we say \emph{complete transmission}:
\begin{definition}
\label{def:complete}
A partial soliton experiences \emph{complete transmission} when total recession, trapping, and incomplete transmission do not occur.
\end{definition}
In the event of complete transmission, the line soliton approaches the RW or DSW, interacts with it, and continues to expand on the far side. As we will demonstrate, complete transmission only occurs in a limited subset of initial conditions. When it does occur, for large $t$ the solution approaches a soliton--mean simple wave described in section~\ref{sec:sw_soln} by the mappings \eqref{eq:f_map}. Numerical simulations with data predicting complete transmission are shown in figures~\ref{fig:lrw1}, \ref{fig:rrw4}, \ref{fig:rdsw2}, and \ref{fig:ldsw3}. In the next section, we will consider the four types of initial conditions shown in figure~\ref{fig:init_cond}. The regions where each type of behaviour is predicted to occur in parameter space for each of the four initial conditions are shown in figure~\ref{fig:transmission region}. Before examining these in detail, we first study partial soliton simple wave--mean flow interactions.

\subsection{Interactions between soliton simple waves and mean flow}
\label{sec:sw_mf_int}

Here we consider in more detail the interaction of a partial soliton simple wave (from section~\ref{sec:ub_zero}) with a RW. Due to the principle of hydrodynamic reciprocity \cite{maiden_hoefer_2018}, we can then extend analogous results to the interaction of a partial soliton simple wave with a DSW. This scenario arises when partial recession occurs, as well as when a transmitted soliton displays multivalued behaviour in the modulation variables.
\par For concreteness, we consider a partial soliton starting to the right of a RW (case (a) in figure~\ref{fig:init_cond}), with initial conditions such that the partial soliton propagates away from the RW. A receding partial soliton starting to the left of a RW experiences total recession and thus does not interact with the RW. In other words, we are studying a Riemann problem with parameters:
\begin{equation}
    \label{eq:int4}
       \ub(x,0) = \begin{cases} 0 & x<0 \\ 1& x>0 \end{cases}, \qquad a(x,0)=\begin{cases} 0 & x<0 \\ a_{\rm R} & x>0 \end{cases}, \qquad
    q(x,0)=\begin{cases} q_* & x<0 \\ q_{\rm R} & x>0 \end{cases},
\end{equation} where $\sqrt{a_{\rm R}}>q_{\rm R}>0$. Following section~\ref{sec:ub_zero}, we determine $q_*$ in order to conserve $R_+$ throughout the solution. This yields
\begin{equation}
    \label{eq:int5}
    q_*^2=2+(q_{\rm R}+\sqrt{a_{\rm R}})^2.
\end{equation}
Note that $q_*$ is always well-defined (real-valued). The right, soliton edge of the partial soliton simple wave has velocity:
\begin{equation*}
    \label{eq:int7}
    U_{\rm s}=1+\frac{a_{\rm R}}{3}-q_{\rm R}^2+\frac{2}{3}\sqrt{a_{\rm R} q_{\rm R}^2} >1.
\end{equation*}
Thus, the soliton edge propagates away from the RW. However, the left, zero edge of the partial soliton simple wave moves left in relation to the mean flow ($U_{\rm z}=1-q_*^2<1$, cf. \eqref{eq:ub_const_sw_3}), interacting with the RW. The two waves intersect at $x=t$, where $\ub(t,t)=R_-(t,t)=1$ and $q(t,t)=\sqrt{a(t,t)}=(q_{\rm R}+\sqrt{a_{\rm R}})/2$. We denote the amplitude at the intersection point as 
\begin{equation}
    \label{eq:new_amp}
a_{\rm i} = \frac{1}{4}(q_{\rm R}+\sqrt{a_{\rm R}})^2.
\end{equation}
 At this point, the characteristic velocities $\lambda_-$ and $\lambda_{\ub}$ and Riemann invariants $R_-$ and $R_{\ub}$ coalesce. 
\begin{figure}
    \centering
    \includegraphics[scale=.25]{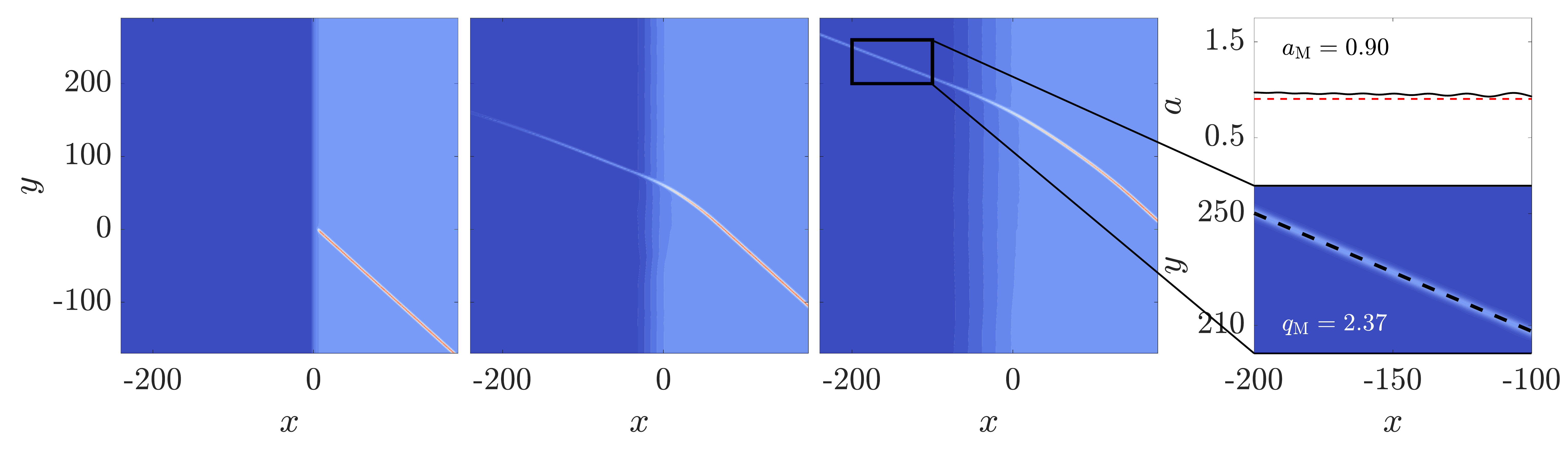}
    \caption{Numerical evolution of the KP equation \eqref{eq:kp} with $a_{\rm R}=4$, $q_{\rm R}=1$, $\ub_{\rm R}=0$, $\ub_{\rm L}=-1$ for $t \in (0,30,80)$. The last panel compares predicted $a_{\rm M}$ and $q_{\rm M}$  \eqref{eq:int15} with $\sigma=-1$ to numerics. This is an example of partial recession with complete transmission.} 
    \label{fig:int1}
\end{figure}
Due to scaling properties of the Riemann problem, the solution must have the self-similar form $R_-=R_-(x/t)$. However, the solution to this problem is non-unique, as there are two possible self-similar solutions satisfying the boundary condition $R_-(t,t)=1$:
\begin{align}
\label{eq:int_soln}
    R_-(x,t) \equiv 1, \quad 0<x \leq t \\
 \label{eq:int10a}
    R_-(x,t) = \frac{x}{t}, \quad 0<x \leq t   
\end{align}
Assuming solution \eqref{eq:int10a}, the definition of the Riemann invariants \eqref{eq:mf3} then requires that $\sqrt{a(x,t)}=q(x,t)$ throughout the RW. We will show that this solution \eqref{eq:int10a} is not possible through a proof by contradiction. If we set $\sqrt{a(x,t)}=q(x,t)$ and insert this into \eqref{eq:mf1}, the amplitude $a$ modulation equation simplifies to
\begin{subequations}
\label{eq:int11}
\begin{equation}
\label{eq:int11b}
    q_t + \frac{q}{3}\ub_x+\left(\ub-\frac{4q^2}{3}\right)q_x-\frac{2}{3}q^2 q_x = 0,
\end{equation}
while the inclination $q$ modulation equation becomes
\begin{equation}
\label{eq:int11c}
    q_t -q\ub_x+\left(\ub-\frac{4q^2}{3}\right)q_x-\frac{2}{3}q^2 q_x = 0.
\end{equation}
\end{subequations}
Subtracting \eqref{eq:int11c} from \eqref{eq:int11b} yields $\frac{4}{3}q\ub_x=0$. For this to be true, either $\ub$ must be constant in $x$, which cannot be the case, or $q\equiv 0$. Since $\ub=x/t$ for $0<x<t$, this implies $a(x,t)\equiv 0$, which cannot be true because $a(t,t)=a_{\rm i}>0$. In short, \eqref{eq:mf1} is not compatible with $\sqrt{a}=q$. This leaves the constant solution \eqref{eq:int_soln} as the only solution. Thus, assumption of constant $R_\pm$ throughout the mean flow still holds even when a partial soliton simple wave interacts with the mean flow.
\par A partial soliton experiencing partial recession can still be transmitted or trapped. Instead of using the initial parameters to determine evolution in the mean flow, we use the parameters at the edge of the RW $(a,q^2)=(a_{\rm i},a_{\rm i})$. The calculation for the soliton amplitude and slope across the RW is the simple wave solution \eqref{eq:mf8} from section~\ref{sec:sw_soln} with $a_0=q_0^2=a_{\rm i}$. The resulting parameters on the left side of the mean flow, if complete transmission occurs, are then
\begin{subequations}
\label{eq:int15}
\begin{align}
    \label{eq:int15a}
    a_{\rm M} &= a_{\rm i}+1+\sigma\sqrt{1+2a_{\rm i}},\\
    \label{eq:int15b}
    q_{\rm M}^2 &= a_{\rm i}+1-\sigma\sqrt{1+2a_{\rm i}},
\end{align}
\end{subequations}
where $\sigma = \pm 1$. Solutions for $a(x,t)$ and $q(x,t)$ within the RW are also described by \eqref{eq:mf11}, again with $q_0^2=a_0=a_{\rm i}$,
\begin{subequations}
\label{eq:int16}
\begin{align}
    \label{eq:int16a}
    a(x,t) &= -\frac{x}{t}+1+a_{\rm i} + \sigma \left[\left(\frac{x}{t}-1 \right)\left(\frac{x}{t}-1-2a_{\rm i} \right)\right]^{1/2}\\
    \label{eq:int16b}
    q^2(x,t) &= -\frac{x}{t}+1+a_{\rm i} - \sigma \left[\left(\frac{x}{t}-1 \right)\left(\frac{x}{t}-1-2a_{\rm i} \right)\right]^{1/2}.
\end{align}\end{subequations}
The earlier definition of $\sigma$ from \eqref{eq:sigma} is not valid here, since $a_0=q_0^2$. Instead, we choose $\sigma$ by appealing to the $\Delta=0$ case presented in section~\ref{sec:ub_zero}. For partial soliton initial conditions given on the right with $\ub$ constant \eqref{eq:ub_zero_2}, the equations for $q(x,t)$ \eqref{eq:qstar1}-\eqref{eq:ub_const_sw_2} imply that $q$ is a monotonically decreasing function of $x$. We expect the same will hold true now with $\Delta \neq 0$. To ensure this, we need to choose $\sigma = -1$. Numerical analysis confirms these predictions. For the specific initial conditions $a_{\rm R}=4$, $q_{\rm R}=1$, $\ub_{\rm R}=0$, $\ub_{\rm L}=-1$, with \eqref{eq:new_amp} inserted into \eqref{eq:int16} and $\sigma=-1$ we predict that left of the RW $a=a_{\rm M}=0.90$ and $q=q_{\rm M}=2.37$. The accuracy of this prediction is confirmed on the fourth panel of figure~\ref{fig:int1}.
\par In summary, the effect of partial recession is that the partial soliton simple wave is ``interrupted" by the RW. The partial soliton simple wave begins to the right of the RW, since here $\lambda_{-}>\lambda_{\ub}$. However, at the right edge of the RW, the ordering of the characteristic velocities $\lambda_-$ and $\lambda_{\ub}$ switches. Throughout the RW and the expanding constant region, $R_{\pm}$ are constant. Then the partial soliton simple wave continues connecting $(a_{\rm M},q_{\rm M})$ to $(0,q_*)$, since for this region $\lambda_{-}<\lambda_{\ub}$. By hydrodynamic reciprocity (see section~\ref{sec:sw_soln}), outside a DSW, $R_{\pm}$ will also be held constant occur when a partial soliton simple wave interacts with a DSW.

\section{Specific cases}
\label{sec:spec}

In this section, we consider the four types of initial conditions displayed in figure~\ref{fig:init_cond}, applying the framework described in section~\ref{sec:gen}. 
\subsection{RW--partial soliton}
\label{sec:lrw}

We first discuss the initial conditions with $\ub_{\rm L}=0$, $\ub_{\rm R}=1$, $a_{\rm R}=a_0$ and $q_{\rm R}=q_0$ given, and $a_{\rm L}=0$, $q_{\rm L}=q_*$.
In this case, from the analysis of a partial soliton with constant mean flow in section~\ref{sec:ub_zero}, we assume that $R_+$ is constant throughout the resulting flow. Only two outcomes are possible: partial recession (leading to complete transmission) and complete transmission.
These initial conditions are shown in panel (a) of figure~\ref{fig:init_cond}, and the corresponding regions of the parameter space $(a_R,q_R^2)$ that give rise to each outcome are shown in panel (a) of figure~\ref{fig:transmission region}. 

\begin{figure}
    \centering
    \includegraphics[scale=.25]{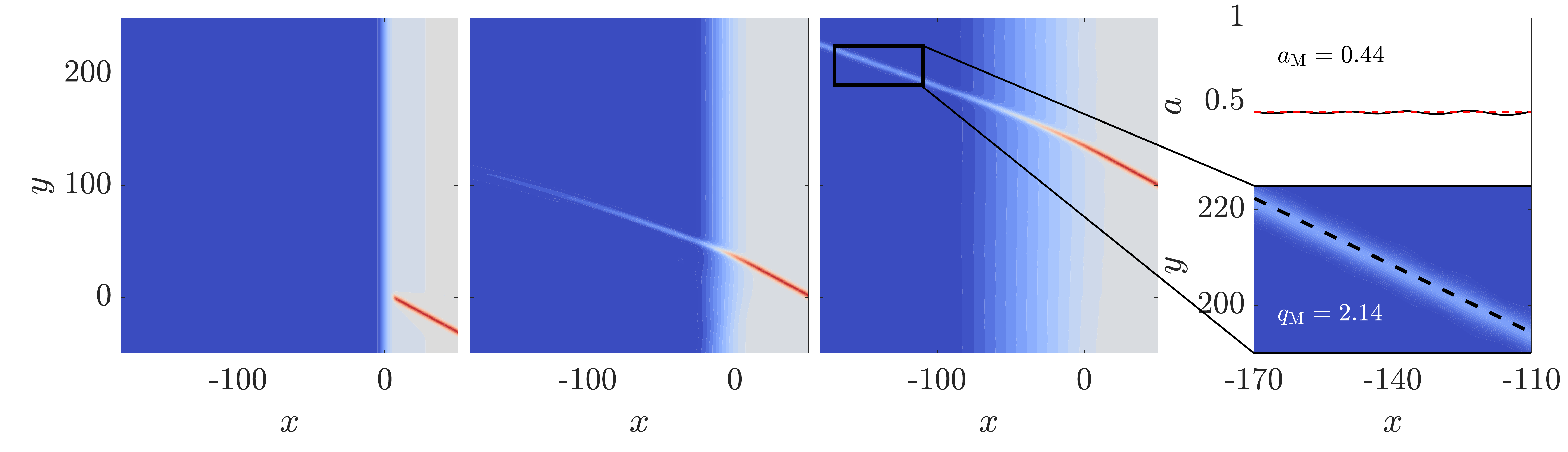}
    \caption{Numerical evolution of the KP equation \eqref{eq:kp} showing complete transmission for the initial conditions (a) in figures \ref{fig:init_cond} and \ref{fig:transmission region}. The initial conditions are $a_{\rm R}=1$, $q_{\rm R}=\sqrt{2}$, $u_R=0$, and $u_L=-1$ on $t\in (0,20,80)$. The right panel compares the parameters of the transmitted soliton with the soliton--mean simple wave prediction \eqref{eq:mf8} and \eqref{eq:f_map_1}.}
    \label{fig:lrw1}
\end{figure}

\paragraph{Partial recession (figure \ref{fig:int1})}
The partial soliton interacts only partially with the RW when $U_{\rm s}>1$ (cf. \eqref{eq:ub_const_sw_2}), implying that
\begin{equation}
    \label{eq:lrw01}
    \sqrt{a_{\rm R}}>q_{\rm R}
\end{equation}
holds for partial recession. This scenario is examined above in section~\ref{sec:sw_mf_int} and depicted in figure~\ref{fig:int1}. 
\paragraph{Complete transmission (figure \ref{fig:lrw1})}
For initial conditions in which partial recession does not occur, the opposite of \eqref{eq:lrw01} must hold, implying that we can fix the sign in \eqref{eq:sigma} for the soliton--mean simple wave to be $\sigma=-1$ throughout the remainder of this section. The transmission condition \eqref{eq:trans_cond} is always met, since $\Delta=1$; trapping never occurs. The mapping in \eqref{eq:mf8} for this case of RW-partial soliton interaction is given in \eqref{eq:f_map_1} yielding
\begin{subequations}
    \label{eq:mf_solution}
\begin{align}
    a_{\rm M} &= \frac{1}{2} \left( a_{\rm R}+ q_{\rm R}^2+2 - \sqrt{(a_{\rm R}+q_{\rm R}^2+2)^2-4 a_{\rm R} q_{\rm R}^2} \right), \\
    q_{\rm M} &= \frac{1}{\sqrt{2}} \left( a_{\rm R}+ q_{\rm R}^2+2 + \sqrt{(a_{\rm R}+q_{\rm R}^2+2)^2-4 a_{\rm R} q_{\rm R}^2} \right)^{1/2}.
\end{align}
\end{subequations}
\par
\begin{figure}
    \centering
     \includegraphics[scale=.25]{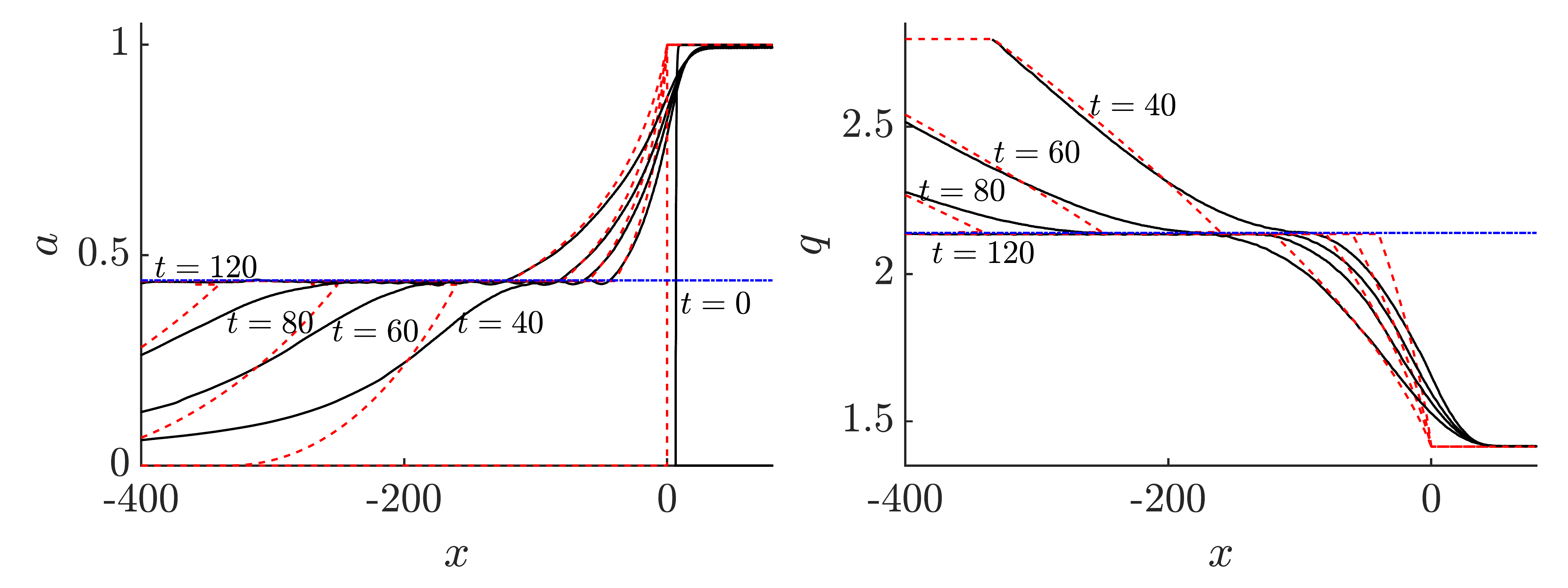}
    \caption{Comparison of analytical results (dashed) from \eqref{eq:lrw_exact} with numerical simulations (solid) for amplitude (left) and slope (right). The numerical simulation the same as figure \ref{fig:lrw1}. A phase shift of $x_0=-20$ is incorporated in the analytical solution. The soliton--mean simple wave result (dash-dotted) from \eqref{eq:mf8} and \eqref{eq:f_map_1} is shown to be the large $t$ limit.}
    \label{fig:lrw2}
\end{figure}

 To the left of the constant region with $a=a_{\rm M}$ and $q=q_{\rm M}$ is a partial soliton simple wave with $R_-$ constant. Incomplete transmission will not occur when $\sqrt{a_{\rm M}}<q_{\rm M}$, which we know from the fact that we fixed $\sigma=-1$ and \eqref{eq:10}. Thus, for this case the soliton always completely transmits. A numerical simulation of complete transmission is shown in figure \ref{fig:lrw1}. The partial soliton parameters on the left side of the RW $(a_{\rm M},q_{\rm M})$ are shown to satisfy the simple wave condition \eqref{eq:mf8} to good accuracy.
\paragraph{Complete transmission exact solution (figure \ref{fig:lrw2})}
When complete transmission occurs, we can calculate explicit modulation solutions for all three parameters. We look for solutions with two simple waves and a constant region of the form:
\begin{subequations}
\label{eq:lrw_exact}
\begin{equation}
    \label{eq:mf17}
    a(x,t) = \begin{cases}
    0 & x<U_{\rm z} t \\
    a_{2}(x,t) & U_{\rm z} t < x < U_{\rm s} t \\
    a_{\rm M} & U_{\rm s} t < x < 0 \\
    a_{3}(x,t) & 0<x < t \\
    a_{\rm R} & t<x
    \end{cases}, \quad
    q(x,t) = \begin{cases}     q_* & x<U_{\rm z} t \\
    q_{2}(x,t) & U_{\rm z} t < x < U_{\rm s} t \\
    q_{\rm M} & U_{\rm s} t < x < 0 \\
    q_{3}(x,t) & 0<x < t \\
    q_{\rm R} & t<x
    \end{cases}.
\end{equation} We already performed the calculations for $a_{3}(x,t)$ and $q_{3}(x,t)$ in \eqref{eq:mf11}:
\begin{align}
    \label{eq:mf21}
    a_{3}(x,t) &= -\frac{x}{t}+\frac{R_{+,\rm R} + R_{-,\rm R}}{2} - \left[\left( \frac{x}{t}-R_{-,\rm R} \right) \left( \frac{x}{t}-R_{+,\rm R} \right)\right]^{1/2}, \\
    \label{eq:mf21b}
    q_{3}^2(x,t) &= -\frac{x}{t}+\frac{R_{+,\rm R} + R_{-,\rm R}}{2} + \left[\left( \frac{x}{t}-R_{-,\rm R} \right) \left( \frac{x}{t}-R_{+,\rm R} \right)\right]^{1/2},
\end{align} 
where $R_{\pm,\rm R} = \ub_{\rm R}+(q_{\rm R}\pm \sqrt{a_{\rm R}})^2$. We use calculations from section~\ref{sec:ub_zero} for the initial condition \eqref{eq:ub_const_sw} to find an explicit formula for $a_{2}(x,t)$ and $q_{2}(x,t)$:
\begin{equation}
    \label{eq:mf22}
    q_{2}(x,t) = \frac{1}{2}\left(q_*^2-3\frac{x}{t}\right)^{1/2}, \qquad \sqrt{a_{2}(x,t)} = q_{\rm M}+\sqrt{a_{\rm M}}-q_{2}(x,t),
\end{equation}
where
\begin{equation}
    \label{eq:mf23}
    U_{\rm s}=\frac{a_{\rm M}}{3}-q_{\rm M}^2+\frac{2}{3}q_{\rm M}\sqrt{a_{\rm M}}, \qquad U_{\rm z}=-q_*^2
\end{equation}
\end{subequations}
A comparison between the above analytical solution \eqref{eq:lrw_exact} and direct numerical simulation is shown in figure \ref{fig:lrw2}. Modulation theory accurately captures the system's behaviour.

\subsection{Partial soliton--RW}
\label{sec:rrw}
Next, we consider initial conditions with $a_{\rm L}=a_0$ and $q_{\rm L}=q_0$ given, $\ub_{\rm L}=0$, $\ub_{\rm R}=1$, $a_{\rm R}=0$, and $q_{\rm R}=q_*$. 
In this case, from the solution to \eqref{eq:ub_zero_3}, $R_-$ should be constant throughout the flow. Four outcomes are possible: complete recession, trapping, incomplete transmission, and complete transmission.
The initial conditions are shown in panel (b) of figure~\ref{fig:init_cond}, and the regions of parameter space that give rise to each outcome are shown in panel (b) in figure~\ref{fig:transmission region}.

\begin{figure}
    \centering
    \includegraphics[]{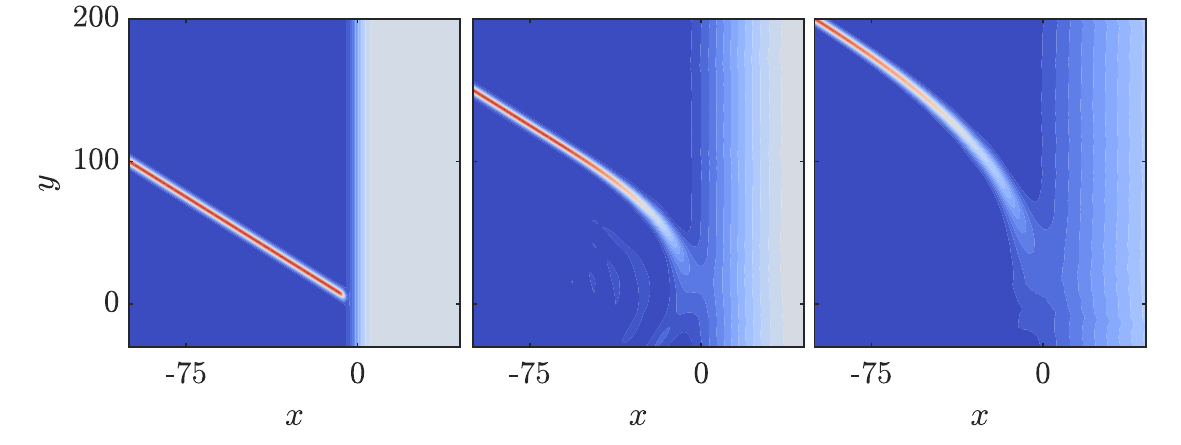}
    \caption{Numerical evolution of the KP equation \eqref{eq:kp} showing total recession for the initial conditions (b) in figures \ref{fig:init_cond} and \ref{fig:transmission region}. The initial parameters are $a_{\rm L}=2$, $q_{\rm L}=1$, $\ub_{\rm R}=1$, and $\ub_{\rm L}=0$ displayed for $t \in (0,30,60)$. This partial soliton  does not interact with the RW.}
    \label{fig:rrw3}
\end{figure}

\begin{figure}
    \centering
    \includegraphics[]{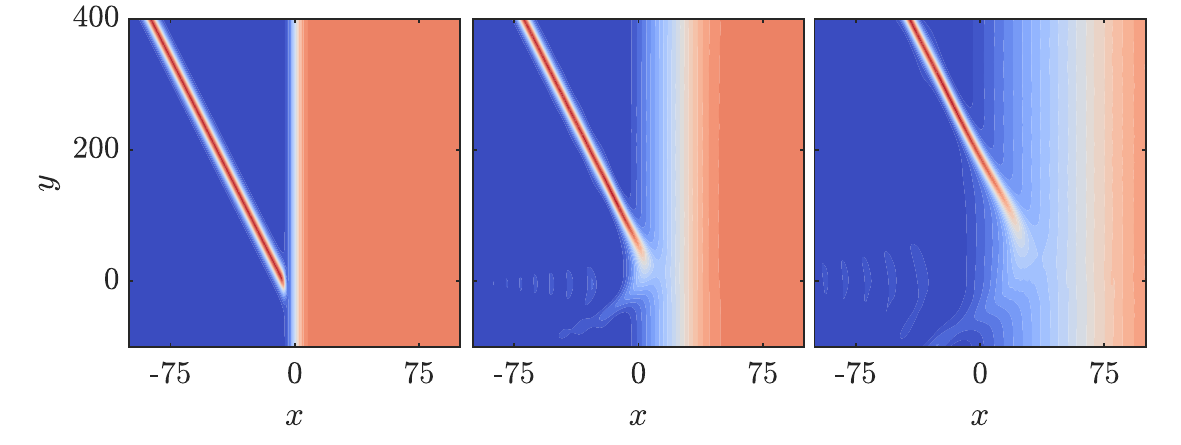}
    \caption{Numerical evolution of the KP equation \eqref{eq:kp} showing trapping for the initial conditions (b) in figures \ref{fig:init_cond} and \ref{fig:transmission region}. The initial parameters are $a_{\rm L}=1.21$, $q_{\rm L}=0.20$, $\ub_{\rm R}=1$, and $\ub_{\rm L}=0$ displayed for $t \in (0,40,100)$. The initial conditions do not satisfy the transmission condition in \eqref{eq:trans_cond}.}
    \label{fig:rrw_trap}
\end{figure}

\begin{figure}
    \centering
    \includegraphics{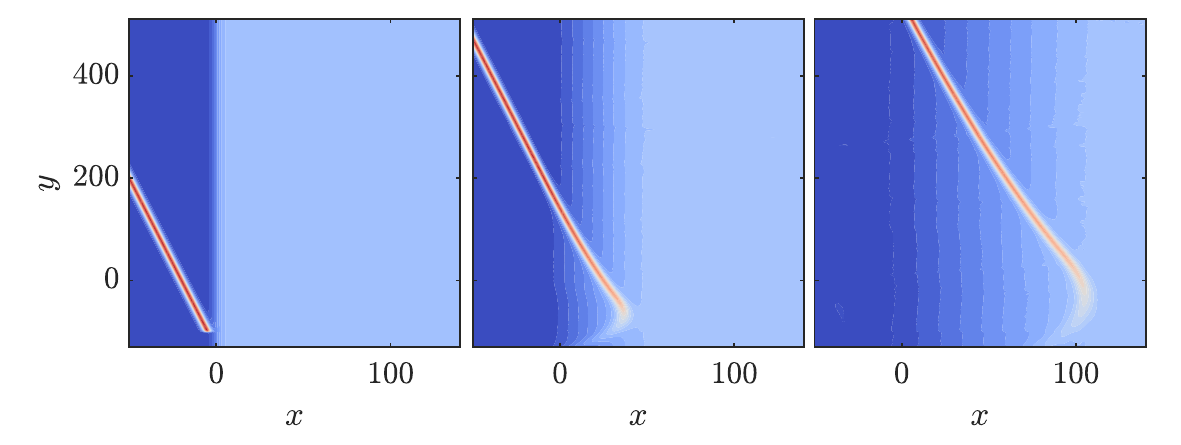}
    \caption{Numerical evolution of the KP equation \eqref{eq:kp} showing incomplete transmission for the initial conditions (b) in figures \ref{fig:init_cond} and \ref{fig:transmission region}. The initial parameters are $a_{\rm L}=3$, $q_{\rm L}=0.15$, $u_R=1$, and $u_L=0$ displayed for $t\in (0,40,100)$. The partial soliton never propagates to the right of the RW.}
    \label{fig:rrw2}
\end{figure}

\begin{figure}
    \centering
    \includegraphics[scale=.25]{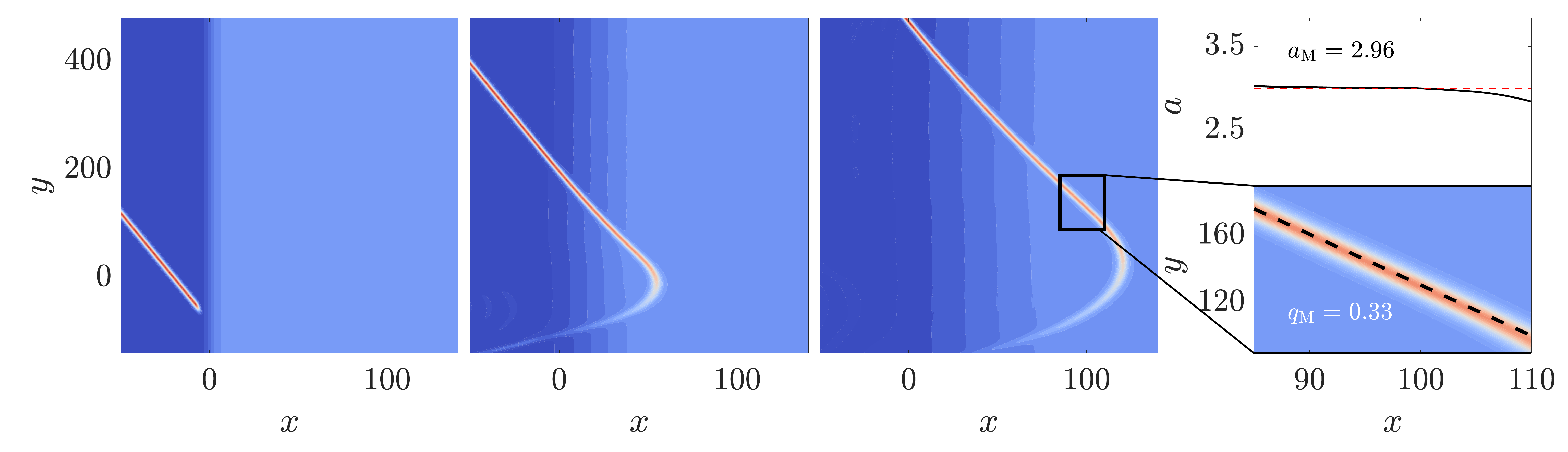}
    \caption{Numerical evolution of the KP equation \eqref{eq:kp} showing complete transmission for the initial conditions (b) in figures \ref{fig:init_cond} and \ref{fig:transmission region}. The initial parameters are $a_{\rm L}=5$, $q_{\rm L}=0.25$, $u_R=1$, and $u_L=0$ displayed for $t\in (0,40,80)$. After transmission the partial soliton bends leftward to completely transmit through the RW again. The right panel compares the parameters of the transmitted soliton with the soliton--mean simple wave prediction \eqref{eq:mf8} and \eqref{eq:f_map_-1}. See figure \ref{fig:rrw5} for a more direct comparison with the modulation solution.}
    \label{fig:rrw4}
\end{figure}

\paragraph{Total recession (figure \ref{fig:rrw3})}
The partial soliton totally recedes from the RW when
\begin{equation}
    \label{eq:rrwa}
    \sqrt{a_{\rm L}}<3q_{\rm L},
\end{equation}
a relatively large portion of the parameter space. A simulation of total recession is shown in figure~\ref{fig:rrw3}. In this case, the partial soliton never interacts with the mean flow. It follows that for a partial soliton under these initial conditions to be transmitted or trapped, $\sqrt{a_{\rm L}}>3q_{\rm L}$, the converse of \eqref{eq:rrwa}, is required. This fixes $a>q^2$ throughout the solution, implying that the sign in \eqref{eq:sigma} for the soliton--mean simple wave must be $\sigma=1$ for the remainder of this section.
\paragraph{Trapping (figure \ref{fig:rrw_trap})}
The transmission condition \eqref{eq:trans_cond} further limits permissible transmitted solutions, as partial solitons can be trapped in the RW. Figure~\ref{fig:rrw_trap} shows an example of this trapped case.
\paragraph{Incomplete transmission (figure \ref{fig:rrw2})}
 If the transmission condition is met and complete recession did not occur, then we can follow \eqref{eq:f_map_-1} to find the parameters of the constant region:
\begin{subequations}
\label{eq:rrw3}
\begin{align}
    \label{eq:rrw3a}
     a_{\rm M} &= \frac{1}{2} \left( a_{\rm L}+ q_{\rm L}^2-2 + \sqrt{(a_{\rm L}+q_{\rm L}^2-2)^2-4 a_{\rm L} q_{\rm L}^2} \right), \\
     \label{eq:rrw3b}
     q_{\rm M} &= \frac{1}{\sqrt{2}} \left( a_{\rm L}+ q_{\rm L}^2-2 - \sqrt{(a_{\rm L}+q_{\rm L}^2-2)^2-4 a_{\rm L} q_{\rm L}^2} \right)^{1/2}.
\end{align}
\end{subequations} If the parameters $(a_{\rm M},q_{\rm M})$ in \eqref{eq:rrw3} do not meet the final condition stated below in \eqref{eq:rrw6}, incomplete transmission occurs. An example of incomplete transmission is shown in figure~\ref{fig:rrw2}. The transmitted soliton never separates from the mean flow.
\paragraph{Complete transmission (figure \ref{fig:rrw4})}
Complete transmission then additionally requires that a constant region develops between the RW and the partial soliton simple wave so that (cf. \eqref{eq:ub_const_sw_2})
\begin{equation}
    \label{eq:rrw6}
    U_{\rm s} = 1 + \frac{a_{\rm M}}{3}-q_{\rm M}^2+\frac{2}{3}\sqrt{a_{\rm M} q_{\rm M}^2}>1,
\end{equation}
which we can write in terms of the incident soliton's parameters using \eqref{eq:rrw3}
\begin{equation}
    \label{eq:rrw7}
    2-(q_{\rm L}-\sqrt{a_{\rm L}})^2+2\sqrt{(a_{\rm L}+q_{\rm L}^2-2)^2-4 a_{\rm L} q_{\rm L}^2}>0. 
\end{equation}
Assuming total recession does not occur (\eqref{eq:rrwa} is not met), complete transmission occurs when both \eqref{eq:trans_cond} and \eqref{eq:rrw7} are met. An example is shown in figure~\ref{fig:rrw4}. For large $t$, the soliton--mean simple wave solution occurs across the RW, as expected.

\begin{figure}
    \centering
    \includegraphics[scale=.25]{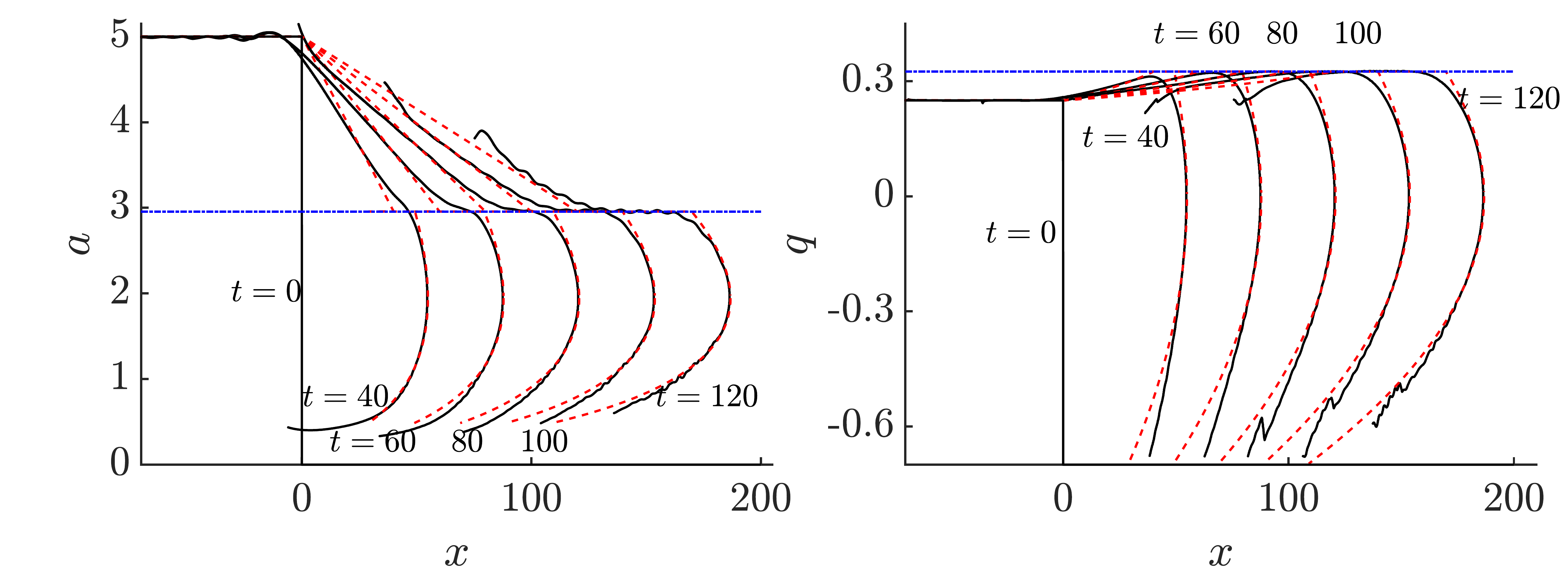}
    \caption{Comparison of analytical results (dashed) from \eqref{eq:rrw_exact} with numerical simulations (solid) for amplitude (left) and slope (right) for complete transmission in initial conditions (b) in \ref{fig:init_cond} and \ref{fig:transmission region}. The numerical simulation is the same as in figure \ref{fig:rrw4}. A phase shift of $x_0=-11$ is incorporated in the analytical solution. The soliton--mean simple wave result (dot-dashed) from \eqref{eq:mf8} and \eqref{eq:f_map_-1} is shown to be the large $t$ limit}
    \label{fig:rrw5}
\end{figure}

\paragraph{Complete transmission exact solution (figure \ref{fig:rrw5})}
We can write down an exact solution in the complete transmission case, which will require utilizing multiple branches of the simple wave solution due to multivalued partial soliton evolution (cf. section~\ref{sec:ub_zero}). Recall that $a_{\rm M}>q_{\rm M}^2$ because $\sigma=1$ \eqref{eq:10}, which from \eqref{eq:qstar2} implies that $q_*<0$. Consequently, the simple wave on the far side of the RW becomes multivalued in $x$. From figure~\ref{fig:rrw4} we see that this is indeed the case. In fact, the solution then curves back around to interact with the RW again, and there we must apply the analysis from Section~\ref{sec:sw_mf_int}. We look for a two branched solution where for $y>V_{\rm f} t$, with $V_{\rm f} = -\frac{2}{3}(q_{\rm M}-\sqrt{a_{\rm M}})$ and $U_{\rm f} = 1+\frac{1}{3}(q_{\rm M}-\sqrt{a_{\rm M}})^2 $ we have
\begin{subequations}
\label{eq:rrw_exact}
\begin{equation}
    \label{eq:rrw_exact_a}
    a^+ (x,t) = \begin{cases}
    a_{\rm L} & x<0 \\
    a_{2}(x,t) & 0 < x < t \\
    a_{\rm M} & t<x<U_{\rm s} t \\
    a_{3}^+(x,t) & U_{\rm s} t < x<U_{\rm f} t
    \end{cases}, \qquad
    q^+(x,t) = \begin{cases}
    q_{\rm L} & x<0 \\
    q_{2}(x,t) & 0 < x < t \\
    q_{\rm M} & t<x<U_{\rm s} t \\
    q_{3}^+(x,t) & U_{\rm s} t <x< U_{\rm f} t
    \end{cases},
\end{equation}
with the characteristic velocity $U_{\rm s}$ defined as in \eqref{eq:rrw6}. For $y<V_{\rm f} t$ we have
\begin{equation}
    \label{eq:rrw_exact_b}
    a^- (x,t) =
    a_{3}^-(x,t), \qquad
    q^-(x,t) =  q_{3}^-(x,t), \quad t<x<U_{\rm f} t,
\end{equation}
 defined below. We can calculate the exact solution within the simple wave just as above in \eqref{eq:mf11}:
\begin{align}
    \label{eq:rrw11}
    a_2(x,t) &= -\frac{x}{t}+\frac{R_{+,{\rm L}} + R_{-,{\rm L}}}{2} - \left[\left( \frac{x}{t}-R_{-,{\rm L}} \right) \left( \frac{x}{t}-R_{+,{\rm L}} \right)\right]^{1/2},\\
    \label{eq:rrw12}
    q_2^2(x,t) &= -\frac{x}{t}+\frac{R_{+,{\rm L}}  + R_{-,L}}{2} + \left[\left( \frac{x}{t}-R_{-,{\rm L}} \right) \left( \frac{x}{t}-R_{+,{\rm L}}  \right)\right]^{1/2},
\end{align}
where $R_{\pm,\rm L} = \ub_{\rm L}+(q_{\rm L} \pm \sqrt{a_{\rm L}})^2$.
The simple wave solution then is
\begin{equation}
    \label{eq:rrw13}
    \begin{split}
    q_3^\pm(x,t) &= \pm \frac{1}{2}\left[(q_{\rm M}-\sqrt{a_{\rm M}})^2+3\left(1-\frac{x}{t}\right)\right]^{1/2}, \\ \sqrt{a_3^\pm(x,t)} &= -q_{\rm M}+\sqrt{a_{\rm M}}+q_3^\pm(x,t).
    \end{split}
\end{equation}\end{subequations}
\par The partial soliton then interacts with the mean flow again at $x=t$, $q_3^-(t,t)=-\sqrt{a_3^-(t,t)} = -\sqrt{a_{\rm i}}$, where $a_{\rm i} = \frac{1}{4}(q_{\rm M}-\sqrt{a_{\rm M}})^2$. Consequently, we also have that $R_+=\ub$ at $x=t$. This is now identical to the partial scenario examined above in section~\ref{sec:sw_mf_int} with the transformation $y \rightarrow -y$ and $R_\pm \rightarrow R_\mp$. The partial soliton, now with parameters $(a,q^2) = (a_{\rm i},a_{\rm i})$, is completely transmitted back through the RW with new parameters defined by \eqref{eq:int15}. At the left side of the RW, a $R_+$-wave continues. 
\par In figure~\ref{fig:rrw5} we show a comparison between direct numerical simulation and the above analytical results \eqref{eq:rrw_exact}. We incorporate a phase shift of $x_0=-11$ to account for higher order effects due to the smooth initial data. The behaviour of the numerical solution is again well-captured by the modulation theory prediction.
\subsection{Partial soliton--DSW}
\label{sec:rdsw}
Now we consider initial conditions where $a_{\rm L}$ and $q_{\rm L}$ are given, $\ub_{\rm L}=1$, $\ub_{\rm R}=0$, $a_{\rm R}=0$, and $q_{\rm R}=q_*$. The mean flow initial conditions will yield a DSW, and $R_-$ is constant throughout the flow following analysis for \eqref{eq:ub_zero_3}. Three outcomes are possible: total recession, incomplete transmission, and complete transmission. The initial conditions are shown in panel (c) of figure~\ref{fig:init_cond}, and the corresponding regions of parameter space for each outcome are shown in panel (c) of figure~\ref{fig:transmission region}.

\paragraph{Total recession}
The partial soliton does not interact with the mean flow when $U_{\rm z}<-1$ (cf. \eqref{eq:ub_const_sw_3}) implying
\begin{equation}
    \label{eq:rdsw01}
    a_{\rm L}-3q_{\rm L}^2-2q_{\rm L}\sqrt{a_{\rm L}}<-2,
\end{equation}
holds for total recession. It follows that for incomplete and complete transmission, the converse of \eqref{eq:rdsw01} must hold.

\begin{figure}
    \centering
    \includegraphics[scale=.25]{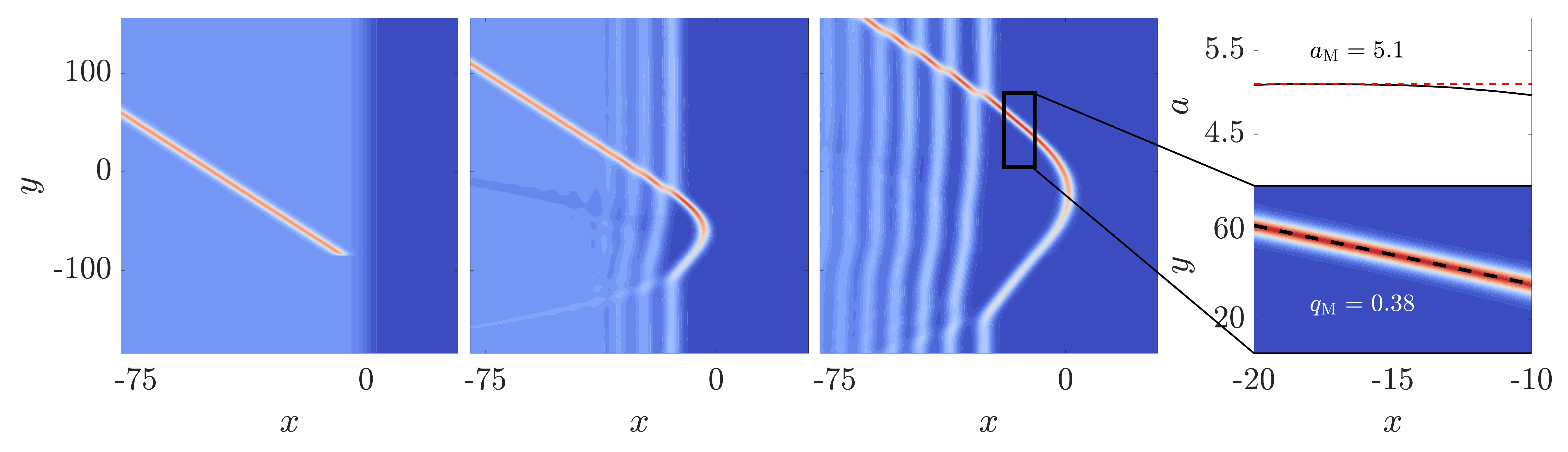}
    \caption{Numerical evolution of the KP equation \eqref{eq:kp} showing complete transmission for the initial conditions (c) in figures \ref{fig:init_cond} and \ref{fig:transmission region}. The initial parameters are $a_{\rm L}=3$, $q_{\rm L}=0.5$, $u_R=-1$, and $u_L=0$ displayed for $t\in (0,20,50)$. Note the zero edge of the partial soliton simple wave bends back to interact with the DSW again. The right panel compares the parameters of the transmitted soliton with the soliton--mean simple wave prediction \eqref{eq:mf8} and \eqref{eq:f_map_1}.}
    \label{fig:rdsw2}
\end{figure}

\paragraph{Incomplete transmission} 
Following \eqref{eq:trans_cond} and \eqref{eq:f_map_1}, we find that a soliton--mean simple wave solution always exists for these initial conditions with parameters
\begin{subequations}
    \label{eq:rdsw2}
\begin{align}
    a_{\rm M} &= \frac{1}{2} \left( a_{\rm L}+ q_{\rm L}^2+2 + \sigma \sqrt{(a_{\rm L}+q_{\rm L}^2+2)^2-4 a_{\rm L} q_{\rm L}^2} \right), \\
    q_{\rm M} &= \frac{1}{\sqrt{2}} \left( a_{\rm L}+ q_{\rm L}^2+2 - \sigma \sqrt{(a_{\rm L}+q_{\rm L}^2+2)^2-4 a_{\rm L} q_{\rm L}^2} \right)^{1/2},
\end{align}
\end{subequations}
with $\sigma$ defined as in \eqref{eq:sigma}. The below condition \eqref{eq:rdsw4} must also be met for complete transmission; if it is not, incomplete transmission occurs and the soliton never separates from the mean flow.
\paragraph{Complete transmission (figure \ref{fig:rdsw2})} The only remaining factor for complete transmission is the magnitude of the partial soliton characteristic velocity on the right side of the DSW $U_{\rm s}>2/3$ (cf. \eqref{eq:ub_const_sw_3}), leading to the condition
\begin{equation}
    \label{eq:rdsw4}
    2-(q_{\rm L}-\sqrt{a_{\rm L}})^2+2\sigma\sqrt{(a_{\rm L}+q_{\rm L}^2+2)^2-4 a_{\rm L} q_{\rm L}^2}>\frac{2}{3}.
\end{equation} 
When the above conditions \eqref{eq:rdsw2} and \eqref{eq:rdsw4} are met, and total recession \eqref{eq:rdsw01} did not occur, complete transmission occurs. Figure~\ref{fig:rdsw2} shows the complete transmission of a partial soliton starting on the left through a DSW. For large $t$, the solution approaches the soliton--mean simple wave from section~\ref{sec:sw_soln}. Once again, the simple wave bends back to interact with the DSW a second time, as described in section~\ref{sec:sw_mf_int}. The secondary interaction occurs at $x/t=2/3$ and is identical to the partial recession case for DSW-soliton initial conditions with the transformation $y \rightarrow -y$ and $R_\pm \rightarrow R_\mp$. In contrast to partial recession for partial soliton-RW data, for this secondary interaction, trapping always occurs. We give details below in section~\ref{sec:ldsw}. 

\subsection{DSW--partial soliton}
\label{sec:ldsw}

When the partial soliton starts to the right of the a DSW, we fix $a_{\rm R}$ and $q_{\rm R}$, $\ub_{\rm L}=1$, $\ub_{\rm R}=0$, $a_{\rm L}=0$, and $q_{\rm L}=q_*$.
From the solution to \eqref{eq:ub_zero_2}, $R_+$ is constant outside the DSW. 
Four outcomes are possible here: partial recession (leading to resonant trapping), trapping (resonant or nonresonant), incomplete transmission, and complete transmission. The initial conditions are shown in panel (d) of figure~\ref{fig:init_cond}, and the corresponding regions of parameter space that give rise to each outcome are shown in panel (d) in figure~\ref{fig:transmission region}.

\paragraph{Partial recession}
The partial soliton only interacts partially with the DSW if $U_{\rm s}>2/3$ (cf. \eqref{eq:ub_const_sw_2}),
 which in our case reads 
\begin{equation}
    \label{eq:rdsw02}
    \frac{a_{\rm R}}{3}-q_{\rm R}^2+\frac{2}{3}q_{\rm R}\sqrt{a_{\rm R}}>\frac{2}{3}.
\end{equation}
When partial recession occurs, the portion of the partial soliton simple wave that interacts with the front edge of the RW is  delimited by the speed  $U=x/t=2/3$,  which corresponds to replacing $a_{\rm R}$ and $q_{\rm R}$ in \eqref{eq:rdsw02} with values such that the inequality becomes an equality.   

We denote the  soliton  parameters at the front edge of the DSW as $(a,q)=(a_{\rm i},q_{\rm i})$. From \eqref{eq:ub_const_sw}, we obtain $q_{\rm i} = q(2t/3,t)=\frac{1}{2}[(q_{\rm R}+\sqrt{a_{\rm R}})^2-2]^{1/2}$, where $q_{\rm i}$ is guaranteed to be real from \eqref{eq:ldsw8b}. We calculate again using \eqref{eq:ub_const_sw}
\begin{equation}
    \label{eq:ldsw20}
    (q_{\rm i}-\sqrt{a_{\rm i}})^2 = \left[(q_{\rm R}+\sqrt{a_{\rm R}})^2-2\right]^{1/2}-(q_{\rm R}+\sqrt{a_{\rm R}}), 
    \end{equation}
from which it is necessary that $(q_{\rm i}-\sqrt{a_{\rm i}})^2\leq 2$ with equality only if $q_{\rm R}+\sqrt{a_{\rm R}}=\sqrt{2}$. Thus, for partial recession the transmission condition \eqref{eq:trans_cond} never holds, i.e. a partial soliton experiencing partial recession from a DSW always remains trapped. Specifically, it experiences resonant trapping, which we will explain below.

\begin{figure}[]
    \centering
    \includegraphics{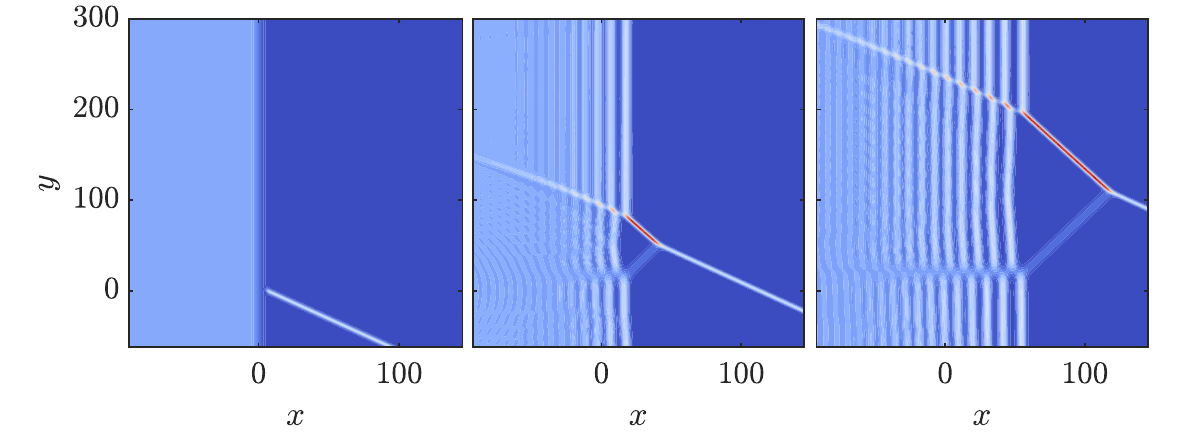}
    \caption{Numerical evolution of the KP equation \eqref{eq:kp} showing resonant trapping for the initial conditions (d) in figures \ref{fig:init_cond} and \ref{fig:transmission region}. The initial parameters $a_{\rm R}=2$, $q_{\rm R}=\sqrt{2}$, $u_R=0$, and $u_L=1$ displayed for $t\in (0,40,100)$. Note the Miles resonant Y-shape forming between the initial partial soliton and the leading edge of the DSW. Both arms of the modulated Y-soliton are trapped by the DSW.}
    \label{fig:ldsw1}
\end{figure}

\paragraph{Resonant trapping (figure \ref{fig:ldsw1})} 
  We now discuss the case where the partial soliton fully interacts with the mean flow, i.e. the partial recession condition \eqref{eq:rdsw02} does not hold. If the transmission condition \eqref{eq:trans_cond} is also not met, trapping occurs, which can take two forms. First, we can have a resonant interaction between the partial soliton and the front edge of the DSW. Recall that the leading edge of a DSW takes the form of a soliton with $a=2$ and $q=0$ for large $t$ (see section~\ref{sec:kdv_red}). From \cite{Biondini_2007}, we find that a resonant interaction occurs between the partial soliton and the DSW leading edge when both of the  following inequalities hold: 
\begin{subequations}
\label{eq:ldsw8}
\begin{align}
    \label{eq:ldsw8a}
    & (q_{\rm R}-\sqrt{a_{\rm R}})^2  < 2, \\
    \label{eq:ldsw8b}
    & (q_{\rm R}+\sqrt{a_{\rm R}})^2  > 2. 
\end{align}\end{subequations}
 Condition \eqref{eq:ldsw8a} is equivalent to the trapping condition \eqref{eq:trans_cond}, and therefore is automatically satisfied if trapping has occurred.  
If \eqref{eq:ldsw8b} also holds, the trapped partial soliton interacts resonantly with the DSW front to form a stem and a second resonant branch between them in a Y-shape. Incidentally, the resonance conditions \eqref{eq:ldsw8} are identical to the conditions that guarantee a complex value in \eqref{eq:mf8} (cf. \eqref{eq:nonneg}). The parameters of the resonant stem are \cite{Biondini_2007}:
\begin{equation}
    \label{eq:ldsw12}
    q_{\rm stem}=\frac{1}{2}(q_{\rm R}+\sqrt{a_{\rm R}}-\sqrt{2}), \qquad \sqrt{a_{\rm stem}} = \frac{1}{2}(q_{\rm R}+\sqrt{a_{\rm R}}+\sqrt{2})
\end{equation} We now use the above parameters in \eqref{eq:mf8} and \eqref{eq:f_map} to verify that trapping does occur. For the parameters \eqref{eq:ldsw12}, the term underneath the square root in the soliton-mean simple wave mapping \eqref{eq:mf8} is zero, since $q_{\rm stem}-\sqrt{a_{\rm stem}}=\sqrt{2}$, implying that $a_{\rm M}=q_{\rm M}^2$ (cf. \eqref{eq:10}).
 The characteristic velocity of the soliton edge of the partial soliton simple wave on the left side of the jump in mean flow then is $U_{\rm s} = 1$ (cf. \eqref{eq:ub_const_sw_2}), equivalent to the velocity of the background flow. Consequently, a partial soliton in this regime cannot establish itself outside the DSW. The other branch of the resonant Y-shape has parameters \cite{Biondini_2007}
 \begin{equation}
     \label{eq:ldsw100}
         q_{\rm res}=\frac{1}{2}(q_{\rm R}-\sqrt{a_{\rm R}}-\sqrt{2}), \quad \sqrt{a_{\rm res}} = \frac{1}{2}(q_{\rm R}-\sqrt{a_{\rm R}}+\sqrt{2}).
 \end{equation} The same analysis holds for $(a_{\rm res},q_{\rm res})$ as for $(a_{\rm stem},q_{\rm stem})$ above, revealing that neither the stem nor the other branch of the resonant Y-soliton will transmit through the DSW. A numerical simulation of this case is shown in figure \ref{fig:ldsw1}.
 \par
\begin{figure}[]
    \centering
    \includegraphics{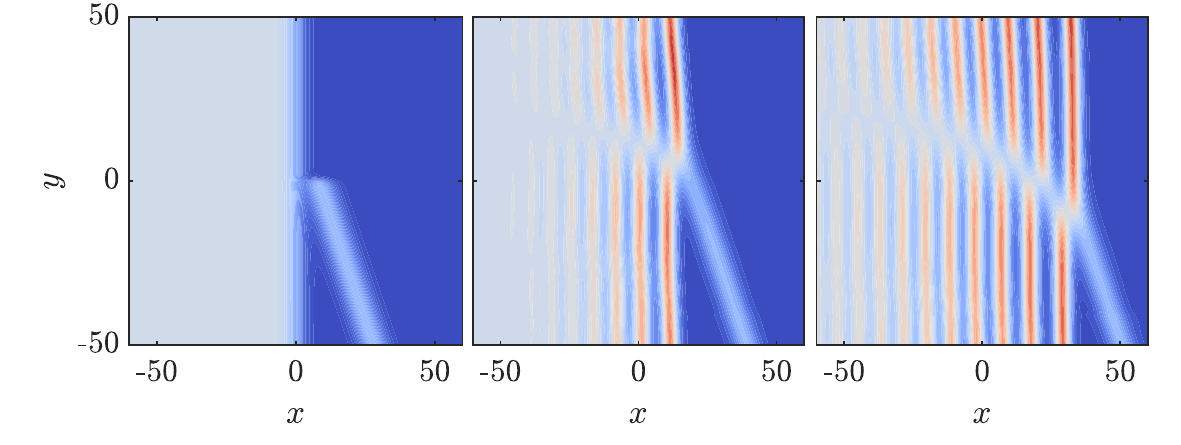}
    \caption{Numerical evolution of the KP equation \eqref{eq:kp} showing nonresonant trapping for the initial conditions (d) in figures \ref{fig:init_cond} and \ref{fig:transmission region}. The initial parameters are $a_{\rm R}=0.8$, $q_{\rm R}=0.4$, $u_R=0$, and $u_L=1$ displayed for $t\in (0,30,60)$. The condition in \eqref{eq:trans_cond} for transmission is not met, and neither is the condition for resonance \eqref{eq:ldsw8b}. Consequently, the partial soliton is trapped and disappears into the DSW.}
    \label{fig:ldsw2}
\end{figure}

 We note that the partial recession condition \eqref{eq:rdsw02} implies that the second resonance condition \eqref{eq:ldsw8b} is met. We already stated that partial recession also implies trapping, which is equivalent to the first resonance condition \eqref{eq:ldsw8a}. Thus, partial recession always leads to resonant trapping.
 \paragraph{Nonresonant trapping (figure \ref{fig:ldsw2})}
 If a partial soliton is trapped per condition \eqref{eq:trans_cond} but does not meet the second condition for resonance \eqref{eq:ldsw8b}, then the second kind of trapping occurs. Trapping without resonance occurs when
 \begin{equation}
    \label{eq:nonresonant trapping}
     (q_{\rm R}+\sqrt{a_{\rm R}})^2<2.
 \end{equation}
 Here the partial soliton is simply absorbed into the DSW, without resonance. This is shown in the numerical simulation of figure~\ref{fig:ldsw2}. Specifically, the interaction between the DSW leading edge soliton and the initial partial soliton is an asymmetric interaction; in fact, from \cite{Biondini_2007}, the regime of nonresonant trapping \eqref{eq:nonresonant trapping} is identical to regime of asymmetric interaction between the DSW leading edge soliton and the initial partial soliton. Inspection of figure~\ref{fig:ldsw2} reveals that the top half of the DSW leading edge has a phase shift to the right, a key feature of asymmetric interactions. 

\paragraph{Incomplete transmission}
When the transmission \eqref{eq:trans_cond} condition is met and partial recession \eqref{eq:rdsw02} did not occur, the partial soliton interacts with the changing mean flow and transmits through it. From \eqref{eq:trans_cond} and \eqref{eq:rdsw02}, we conclude that if transmission occurs, $q_R^2>a_R$, implying that $\sigma=-1$ \eqref{eq:sigma}. The parameters $a_{\rm M}$ and $q_{\rm M}$ are well-defined and are given by (cf. \eqref{eq:mf8} and \eqref{eq:f_map_-1})
\begin{subequations}
    \label{eq:lsdw17}
\begin{align}
    a_{\rm M} &= \frac{1}{2} \left( a_{\rm R}+ q_{\rm R}^2-2 - \sqrt{(a_{\rm R}+q_{\rm R}^2-2)^2-4 a_{\rm R} q_{\rm R}^2} \right) ,\\
    q_{\rm M} &= \frac{1}{\sqrt{2}} \left( a_{\rm R}+ q_{\rm R}^2-2 + \sqrt{(a_{\rm R}+q_{\rm R}^2-2)^2-4 a_{\rm R} q_{\rm R}^2} \right)^{1/2}.
\end{align}
\end{subequations}
Incomplete transmission then occurs if the condition for complete transmission given below \eqref{eq:ldsw19} is not met.

\begin{figure}[]
    \centering
    \includegraphics[scale=.25]{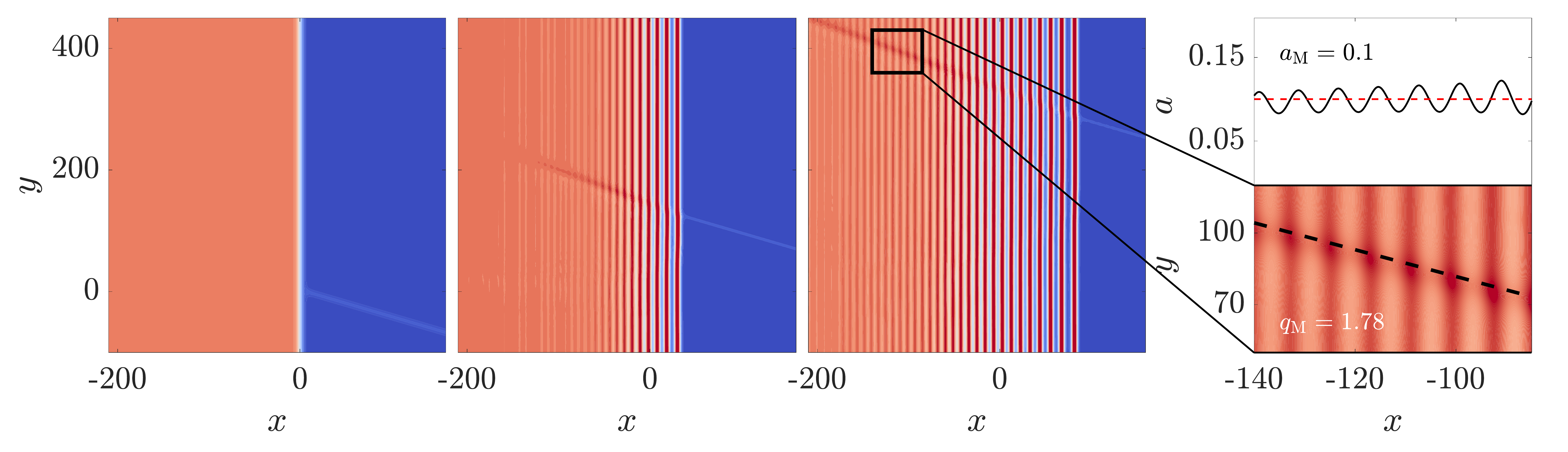}
    \caption{Numerical evolution of the KP equation \eqref{eq:kp} showing complete transmission for the initial conditions (d) in figures \ref{fig:init_cond} and \ref{fig:transmission region}. The initial parameters are $a_{\rm R}=0.06$, $q_{\rm R}=2.28$, $u_R=0$, and $u_L=1$ displayed for $t\in (0,60,140)$. The background DSW computed from another simulation is subtracted from the numerical simulation in order to render visible the growing line soliton. The right panel compares the parameters of the transmitted soliton with the soliton--mean simple wave prediction \eqref{eq:mf8} and \eqref{eq:f_map_-1}.}
    \label{fig:ldsw3}
\end{figure}

\paragraph{Complete transmission  (figure \ref{fig:ldsw3})}
For complete transmission, we require that on the left side of the mean flow $U_{\rm s} < -1$, which implies
\begin{equation}
    \label{eq:ldsw19}
    5< (q_{\rm R}-\sqrt{a_{\rm R}})^2+2\sqrt{(a_{\rm R}+q_{\rm R}^2-2)^2-4 a_{\rm R} q_{\rm R}^2},
\end{equation}
otherwise incomplete transmission occurs.
If the above condition \eqref{eq:ldsw19} holds, complete transmission occurs.  An example of complete transmission is shown in figure~\ref{fig:ldsw3}. This soliton has very small amplitude $a_{\rm R}$, but the large $q_{\rm R}$ ensures transmission. Even in this extreme regime, the solution approaches the soliton--mean simple wave solution from section~\ref{sec:sw_soln}. We also remark here that the transmitted soliton interaction with the DSW leading edge is classified as an ordinary interaction as defined in \cite{kodama_book,Biondini_2007}. In fact, the regime for an ordinary interaction from \cite{Biondini_2007} is identical to the transmission condition \eqref{eq:trans_cond} for a soliton starting to the right of a DSW. Looking closely at figure~\ref{fig:ldsw3} reveals a leftward phase shift for a DSW leading edge above the initial soliton, an indicator of an ordinary interaction \cite{Biondini_2007}.

\section{Discussion and conclusion}
\label{sec:concl}

The main goal of this paper was to investigate the interaction of an oblique soliton with a dispersive hydrodynamic mean flow. 
In particular, we showed that, even though Whitham modulation theory is not as often used for (2+1)-dimensional equations, 
it provides a tractable and reliable analytical approach for examining the evolution of line solitons of the KP equation, even under step-like initial conditions for the mean flow. 
In a previous work \cite{ryskamp_2020}, the $x$-independent reduction of the KP soliton modulation equations was considered. 
In this study, in contrast, we primarily analyzed the $y$-independent reduction. Specifically, by diagonalizing the $y$-independent KPII soliton--mean flow modulation system \eqref{eq:mf1}, we were able to calculate invariant quantities that constrain admissible modulated line solitons through a changing mean flow. As a consequence of the loss of strict hyperbolicity, there are two distinct possible combinations of soliton parameters across the mean flow. We determine a unique solution by appealing to consistency with the evolution of the parameters within a RW. This finding implies that, in general, $\sigma = \sgn{a-q^2}$ is constant within a mean flow, i.e. the parameters cannot cross the plane $a=q^2$ where strict hyperbolicity is lost. Based on this restriction, we determine unique mappings for the soliton--mean simple wave, which are shown to be large $t$ attractors for the partial soliton--mean flow initial conditions in numerical simulations. As may be expected, the mapping for a transmitted soliton from a higher mean flow to a lower mean flow is an exact inverse of the mapping for a transmitted soliton from the lower to higher mean flow.
\par A key result of this paper is the transmission condition \eqref{eq:trans_cond}, which generalizes the KdV transmission condition previously calculated in \cite{ablowitz_cole_2018,maiden_hoefer_2018}. This relation implies that while a line soliton can always transmit through a downward jump in the mean, a line soliton only transmits through an upward jump if $(q-\sqrt{a})^2>2\Delta$, where $\Delta$ is the size of the jump. The transmission condition \eqref{eq:trans_cond} also predicts a number of novel behaviours not present in the (1+1)-dimensional setting, such as backward transmission: any nonzero $q$ allows a soliton to transmit through a RW starting from the right, whereas a ($q=0$) KdV soliton does not interact with a RW and is trapped by a DSW if its amplitude is less than twice the jump in the mean. Another new feature is that certain inclinations can prevent transmission of even a large amplitude soliton through a RW, such as choosing $q=\sqrt{a}$, while a very small amplitude soliton can pass through both RWs and DSWs from either direction with sufficiently large inclination (see figure~\ref{fig:ldsw3} for an example).  
\par 
We also discovered a remarkable connection between the various outcomes for transmission of a soliton through a DSW  
and the classifications of the interactions between two line solitons \cite{Biondini_2007}. 
Indeed, the three possible outcomes for a soliton incident on a DSW match precisely the three classes of exact solutions asymptoting to two line solitons as $y \to \pm \infty$. 
Specifically, the 
soliton is transmitted through a DSW (cf. \eqref{eq:trans_cond}) if and only if an ordinary interaction occurs between the line soliton on the right and the soliton at the leading edge of the DSW. 
Otherwise, a resonant trapping 
(which corresponds to the case in which the mapping \eqref{eq:mf8} yields no acceptable solutions) occurs
if and only if 
a resonant interaction occurs between the line soliton and the leading edge of the DSW (cf.~\eqref{eq:nonneg} and \eqref{eq:ldsw8}). 
Finally, a nonresonant trapping
(which corresponds to the case in which the mapping \eqref{eq:mf8} gives a negative value for $a_1$ or $q_1^2$)
occurs if and only if
an asymmetric interaction occurs between the line soliton and the leading edge of the DSW (cf.~\eqref{eq:nonneg} and \eqref{eq:nonresonant trapping}). 
In other words, the conditions giving rise to each soliton--mean flow interaction scenario correspond precisely to the conditions 
that give rise to ordinary, resonant and asymmetric interactions between the oblique soliton and the soliton at the leading edge of the DSW in \cite{Biondini_2007}.
We also mention that, to the best of our knowledge, this is the first work to report the generation of 
a resonant soliton interaction from the time evolution of essentially non-solitonic initial conditions
(where we use the term ``essentially non-solitonic initial conditions'' to denote initial conditions that are not 
simply a linear or nonlinear superposition of one or more infinite line solitons).
\par 
By piecing together simple wave solutions of the soliton--mean modulation equations, we are also able to predict the dynamic behaviour of a partial line soliton incident upon step initial conditions. Despite regions where the partial soliton becomes multivalued in $x$---another consequence of nonstrict hyperbolicity---it was found that through a fully two-dimensional spatial regularization, a well-defined solution is obtained. For completely transmitting multivalued partial solitons, the soliton bends back around to interact with the mean flow a second time. Modulation theory also predicts other novel behaviours, such as incomplete transmission and partial recession. In general, the complete transmission regions of parameters for the partial soliton--mean flow problem are relatively small (see figure~\ref{fig:transmission region}). In cases of complete transmission, modulation theory yields quantitatively accurate results for both the evolution within a RW, as well as modulation parameters outside the DSW or RW.
\par
Even though in this work we used the $y$-independent modulation system to study the 
the soliton--mean flow problem, we found that there is 
a scenario where it was necessary to appeal to the full modulation system, both to justify a multivalued-in-$x$ solution and to determine its regime of validity. 
A natural continuation of this work is therefore to consider initial conditions that are not limited to one-dimensional reductions. A well-known example of such a solution is the Miles resonant (Y-shaped) soliton. A natural question is therefore how one can study such a solution using modulation theory.
Another extension of our work will be to consider mean flows that are fully two-dimensional. 
Finally, a third area of further research is to examine the interaction of waves besides solitons with mean flows, such as small-amplitude linear waves, cnoidal waves, or DSWs.

\section*{Acknowledgements}
The work of MAH and SR was supported by NSF grant DMS-1816934. The work of GB was
supported by NSF grant DMS-2009487.  Authors thank the Fields
Institute Focus Program on Nonlinear Dispersive Partial Differential
Equations and Inverse Scattering in the summer of 2017 where this
research was initiated.

\appendix

\section{Equivalence of simple waves}
\label{sec:eq_simp_waves}
In this appendix, we will demonstrate the validity of the multivalued simple wave solution \eqref{eq:ub_const_sw_f} to the partial soliton initial conditions \eqref{eq:ub_zero_3} with $\ub \equiv const.$ and $0<q_0<\sqrt{a_0}$. From \eqref{eq:qstar2} and section~\ref{sec:ub_zero}, $q_*<0$ and the solution must become multivalued in $x$. To examine this, we study the $x$-independent soliton modulation equations \cite{biondini2019integrability,ryskamp_2020,neu_singular_2015},
\begin{equation}
  \label{eq:x-independent}
  \begin{bmatrix}
    a \\ q
  \end{bmatrix}_t +
  \begin{bmatrix}
    2q & \frac{4}{3}a \\
    \frac{1}{3} & 2q
  \end{bmatrix}
  \begin{bmatrix}
    a \\ q
  \end{bmatrix}_y
  = 0 ,
\end{equation}
which can be rewritten in diagonalized form as
    \begin{equation}\label{eq:RI_y}
        r_\pm = q \pm \sqrt{a}, \quad V_\pm = 2q\pm\frac{2}{3}\sqrt{a}=\frac{4}{3}r_\pm + \frac{2}{3}r_\mp,  \quad \frac{\partial r_\pm}{\partial t} + V_\pm \frac{\partial r_\pm}{\partial y} = 0.
    \end{equation}
The Riemann problem \eqref{eq:ub_zero_3} in $x$ is equivalently formulated as a Riemann problem in $y$
\begin{equation}
    \label{eq:ub_zero_4}
       a(x,0)=\begin{cases} 0 & y<0 \\ a_0 & y>0 \end{cases}, \quad
    q(x,0)=\begin{cases} q_* & y<0 \\ q_0 & y>0 \end{cases}.
\end{equation}
Since the $a=0$ vacuum region is below the nonzero region and $V_-<V_+$ for $a_0>0$, $r_-$ will first propagate into the vacuum region and determine $q_*$ by \eqref{eq:qstar2}. This ensures that $r_-$ is constant throughout the solution, and the emerging simple wave has $r_+$ changing. For this problem in $y$ there is no loss of strict hyperbolicity or multivalued behaviour, as long as $a_0>0$. The characteristic velocities on either side of the simple wave are
\begin{equation}
    \label{eq:ub_zero_5}
    V_{\rm s} = 2q_0+\frac{2}{3}\sqrt{a_0}, \qquad V_{\rm z} = 2q_*,
\end{equation}
where $V_{\rm s}$ and $V_{\rm z}$ represent the velocities of the soliton and zero edges of the simple wave, respectively. Since $q$ is defined as the slope $\tan{\varphi}=-x/y$ (see figure~\ref{fig:soliton}), we can use the transformations $V_{\rm s}=-U_{\rm s}/q_0$ and $V_{\rm z}=-U_{\rm s}/q_*$ and add the appropriate soliton speed $c=\ub+a/3+q^2$ (see \eqref{eq:soli}) to convert characteristic velocities in $y$ to characteristic velocities in $x$. This yields
\begin{equation}
    \label{eq:app_4}
    U_{\rm s} = \ub+\frac{a_0}{3}-q_0^2-\frac{2}{3}q_0\sqrt{a_0}, \qquad U_{\rm z} = \ub-q_*^2.
\end{equation}
The above characteristic velocities \eqref{eq:app_4} agree with \eqref{eq:ub_zero_3}, which shows that the edge velocities of the partial soliton simple wave are equivalent in $x$ and $y$. It remains to show that the evolution of $r_+$ within the simple wave is the same. Solving for the $r_+$-simple wave solution in $y$ yields
\begin{equation}
    \label{eq:app_5}
    r_+(y,t) = \begin{cases}
    q_0+\sqrt{a_0} & V_{\rm s} t < y \\
    \frac{1}{4}(3y/t-2r_-) & V_{\rm z} t<y<V_{\rm s} t\\
    q_*=q_0-\sqrt{a_0} & y< V_{\rm z} t
    \end{cases}.
\end{equation}
We will now solve the original problem \eqref{eq:ub_zero_3} in a moving reference frame in $x$ and show that the results are identical. The reference frame will travel with the velocity of the soliton:
\begin{equation}
    \label{eq:soli_speed}
    \tilde x = x-ct, \quad c = \frac{a}{3}+q^2 = \frac{1}{3} ( r_+^2 + r_+ r_- + r_-^2).
\end{equation}
Setting $r_+=x/t=\Lambda_+$ \eqref{eq:red_mean_flow} in the simple wave region and solving for $r_+$ gives
\begin{equation}
    \label{eq:app_8}
    r_+(\tilde x,t) = \frac{1}{4}\left(-3r_- \pm \sqrt{r_\pm^2-24\tilde x/t}\right).
\end{equation}
Note that the multivalued nature of solutions in $x$ reappears here. Once again using the change of variables $\tilde x = -y q = -y (r_+ + r_-)/2$ and simplifying yields only one solution
\begin{equation}
    \label{eq:app_9}
    r_+(y,t) = \frac{1}{4}(3y/t-2r_-)
\end{equation} for both the positive and negative square root cases in \eqref{eq:app_8}. This agrees with the solution in \eqref{eq:app_5}. Thus, solving the problem \eqref{eq:ub_zero_3} using $y$-independent modulation equations is equivalent to solving the problem \eqref{eq:ub_zero_4} using $x$-independent modulation equations.
\par In order to determine the location of the branch point $q=0$ in $y$, note that \eqref{eq:RI_y} implies that $q=(r_+ + r_-)/2$. Thus, $q=0$ when $r_+(x,t)=-r_-$ in \eqref{eq:app_5}. Solving for $y/t$ gives
    \begin{equation}
        \label{eq:zero_speed_in_y}
        \frac{y}{t}=-\frac{2}{3}r_-=-\frac{2}{3}q_*.
    \end{equation}

\bibliography{kp_soli_mean_flow.bib}
\end{document}